\documentclass{aa}  

\usepackage{graphicx}
\usepackage{txfonts}
\usepackage{xcolor}
\begin{document} 
\title{Dynamics of intermediate mass black holes in globular clusters
}
   \subtitle{Wander radius and anisotropy profiles}
 \titlerunning{IMBHs in GCs, the effect on velocity dispersion}
\author{Pierfrancesco Di Cintio
          \inst{1,2,3,4}
          \and
          Mario Pasquato
          \inst{5,6}
          \and
        Luca Barbieri
          \inst{2,3,4}
        \and
        Alessandro A. Trani
          \inst{7,8}  
        \and
        Ugo N. di Carlo
          \inst{9}
          }
   \institute{
        Institute of Complex Systems - National Council of Research (ISC-CNR), via della Lastruccia 10, I--50019 Sesto Fiorentino, Italy\\
    \email{pierfrancesco.dicintio@cnr.it}\\    
     \and
         Physics and Astronomy Department, University of Firenze, via G. Sansone 1, I--50019 Sesto Fiorentino, Italy\\
     \and
     INAF, Osservatorio Astrofisico di Arcetri, Largo Enrico Fermi, 5, I-50125 Firenze, Italy\\
     \and      
     INFN - Sezione di Firenze, via G. Sansone 1, I--50019 Sesto Fiorentino, Italy\\
     \and
    Physics and Astronomy Department Galileo Galilei, University of Padova, Vicolo dell'Osservatorio 3, I--35122, Padova, Italy \\
    \and
 Département de Physique, Université de Montréal, Montreal, Quebec H3T 1J4, Canada\\
         \and
     The University of Tokyo, Earth Science and Astronomy Department, 7-3-1 Hongo, Bunkyo-ku, Tokyo 113-0033, Japan\\
     \and
     Okinawa Institute of Science and Technology Graduate University
     1919-1 Tancha, Onna-son, Kunigami-gun 904-0495
      Okinawa, Japan\\
     \and
     McWilliams Center for Cosmology and Department of Physics, Carnegie Mellon University, Pittsburgh, PA 15213, USA
             }
   \date{Received September 15, 1996; accepted March 16, 1997}
  \abstract 
{We recently introduced a new method for simulating collisional gravitational $N$-body systems with approximately linear time scaling with $N$. Our method is based on the Multi-Particle Collision (MPC) scheme, previously applied in Fluid Dynamics and Plasma Physics. We are able to simulate globular clusters with a realistic number of stellar particles (at least up to several times $10^6$) on a standard workstation.}{We simulate clusters hosting an intermediate mass black hole (IMBH), probing a broad range of BH-cluster and BH--average-star mass ratios, unrestricted by the computational constraints that affect direct $N$-body codes.} {We set up a grid of hybrid particle-in-cell--multi-particle collision (MPC) $N$-body simulations using our implementation of the MPC method, MPCDSS. We use either single mass models or models with a Salpeter mass function (a single power-law with exponent $-2.35$), with the IMBH initially sitting at the centre. The force exerted by and on the IMBH is evaluated with a direct sum scheme with or without softening. For all simulations we measure the evolution of the Lagrangian radii and core density and velocity dispersion over time. In addition, we also measure the evolution of the velocity anisotropy profiles.}
{We find that models with an IMBH undergo core collapse at earlier times, the larger the IMBH mass the shallower, with an approximately constant central density at core collapse. The presence of an IMBH tends to lower the central velocity dispersion. These results hold independently of the mass function of the model. For the models with Salpeter MF we observe that equipartition of kinetic energies is never achieved, even long after core collapse. Orbital anisotropy at large radii appears driven by energetic escapers on radial orbits, triggered by strong collisions with the IMBH in the core. We measure the wander radius, i.e. the distance of the IMBH from the centre of mass of the parent system over time, finding that its distribution has positive kurtosis.}{Among the results we obtained, which mostly confirm or extend previously known trends that had been established over the range of parameters accessible to direct N-body simulations, we underline that the leptokurtic nature of the IMBH wander radius distribution might lead to IMBHs presenting as off-centre more frequently than expected, with implications on observational IMBH detection.}
%%%%%%%%%%%%%%%%%%%%%%%%%%%%%%%%%%%%%%%%%%%%%%%%%%%%%%%%%%%%%%%%%%%%%%
   \keywords{(Galaxy:) globular clusters: general - methods: numerical}
   \maketitle
%-------------------------------------------------------------------
\section{Introduction}
A plausible mechanism for super-massive black hole (SMBH) seeding is required to explain the observation of quasars at high redshift \citep[][]{2020ARA&A..58...27I, 2022MNRAS.509.1885P}. Early seeding would rely on pristine gas and is speculated to take place either through direct collapse \citep[][]{1994ApJ...432...52L, 2006MNRAS.371.1813L, 2008MNRAS.383.1079V} or population III stars \citep[][]{1984ApJ...277..445C, 2006ApJ...652....6Y, 2015ComAC...2....3G}. Later or continuous seeding would instead happen through dynamically mediated gravitational runaway scenarios in dense environments \citep[][]{2002MNRAS.330..232C,2001ApJ...562L..19E,2004Natur.428..724P,2022arXiv220811894L}. SMBH seeds should be detectable today as intermediate-mass black holes (IMBHs) in dense stellar systems such as star clusters \citep[][]{2020ARA&A..58..257G,2021MNRAS.507.5132D,2021MNRAS.501.5257R,2023MNRAS.tmp..730R}, especially if the second scenario is prevalent, modulo expulsion from the host system via gravitational wave recoil kicks \citep[see e.g.][]{2008ApJ...686..829H,2022arXiv221116523W}. Quantitatively addressing the seeding mechanism requires us to constrain the fraction of ``rogue'' IMBHs \citep[i.e. not associated to a host star cluster, which may still be detectable by other means, e.g.][]{2018MNRAS.480.4684B}, that requires correctly modelling IMBH ejection from the parent cluster.
Moreover, when looking for electromagnetic signatures of accretion \citep[e.g.][]{2018ApJ...862...16T} it is crucial to have a good estimate of the wander radius of an IMBH within its host star cluster. An underestimate may lead us to exclude off-centre radio sources which could be potential IMBHs.\\ Finally, when an IMBH claim is made based on radial velocity signatures, as in the case of the IMBH in the Leo I dwarf spheroidal \citep[][]{2021ApJ...921..107B}, velocity dispersion anisotropy could be an important source of confusion \citep[][]{2016IAUS..312..197Z}, as strongly radially anisotropic systems could be compatible with a massive central object as well as with radially biased initial conditions. \\
Estimating the probability of IMBH expulsion and the wander radius, and tracing the evolution of anisotropy in the presence of an IMBH are three applications in which the recently introduced MPCDSS code \citep[][]{2021A&A...649A..24D, 2022A&A...659A..19D} becomes competitive in terms of realism with direct $N$-body codes and other approximate approaches, as we argue in the following.\\
\indent Direct $N$-body simulations of stellar systems are often perceived as more realistic (e.g. see \citealt{2000ApJ...535..759T,2001MNRAS.325.1323B,2005MNRAS.363..293H,2008ApJ...685..247B,2016MNRAS.461.1023B,2020MNRAS.491.2413W}) than other approaches such as for example the Monte Carlo methods (e.g. see \citealt{2001A&A...375..711F,2006MNRAS.371..484G,2013MNRAS.429.1221H,2013MNRAS.431.2184G,2014MNRAS.443.3513S,2015MNRAS.446.3150V,2019MNRAS.483.1523S,2020MNRAS.499.4646A}), especially when collisional dynamics is involved (see the discussions in \citealt{2008MNRAS.383....2K,2011BASI...39...69H,2021arXiv210508067K}).\\
However star clusters that are both in a collisional dynamic regime and contain $N>10^6$ particles are a common occurrence, even in our Galaxy \citep[see the discussion in][]{2021A&A...649A..24D}. Direct $N$-body models of these star clusters cannot be simulated using a one-to-one star to stellar-particle ratio, due to the computational constraints of the method. This greatly reduces the faithfulness of direct $N$-body simulations.\\
\indent When dealing with IMBH hosting systems, this limitation can be recast in terms of two dimensionless ratios that are important for determining the dynamical evolution of the system: the ratio of IMBH mass to the average stellar mass in the system, 
\begin{equation}
\mu \equiv M_{\rm IMBH}/\langle m \rangle, 
\end{equation}
and the ratio of IMBH mass to the total mass in the system 
\begin{equation}
\alpha\equiv M_{\rm IMBH}/M.
\end{equation}
We can better appreciate the role of these two ratios by way of a very simplified example, where the star cluster has a typical size $R$, is virialized, and equipartition of kinetic energies holds between the IMBH and the surrounding stars with velocity dispersion $\sigma$. With these assumptions, the radius of the sphere of influence of the IMBH (i.e. the radius below which the BH potential $\Phi_{\rm IMBH}=-GM_{\rm IMBH}/r$ dominates over the contribution of the stellar component, e.g. see \citealt{1972ApJ...178..371P}; see also \citealt{2004cbhg.symp..263M}) is 
\begin{equation}\label{rinflu}
    r_{\rm inf} = \frac{GM_{\rm IMBH}}{\sigma^2}  \approx  R \frac{M_{\rm IMBH}}{M}=R\alpha,
\end{equation} 
while the so called wander radius (the typical distance at which the IMBH is found from the host centre of mass, see \citealt{1976ApJ...209..214B,2011MNRAS.418.1308B}) works out to
\begin{equation}\label{rwander}
    r_{\rm wan} = R \sqrt{\frac{\langle m \rangle}{M_{\rm IMBH}}}=R\mu^{-1/2}.
\end{equation}
%%%%%%%%%%%%%%%%%%%%%%%%%%%%%%%%%%%%%%%%%%%%%%%%%%%
\begin{figure}
    \centering
    \includegraphics[width=\columnwidth]{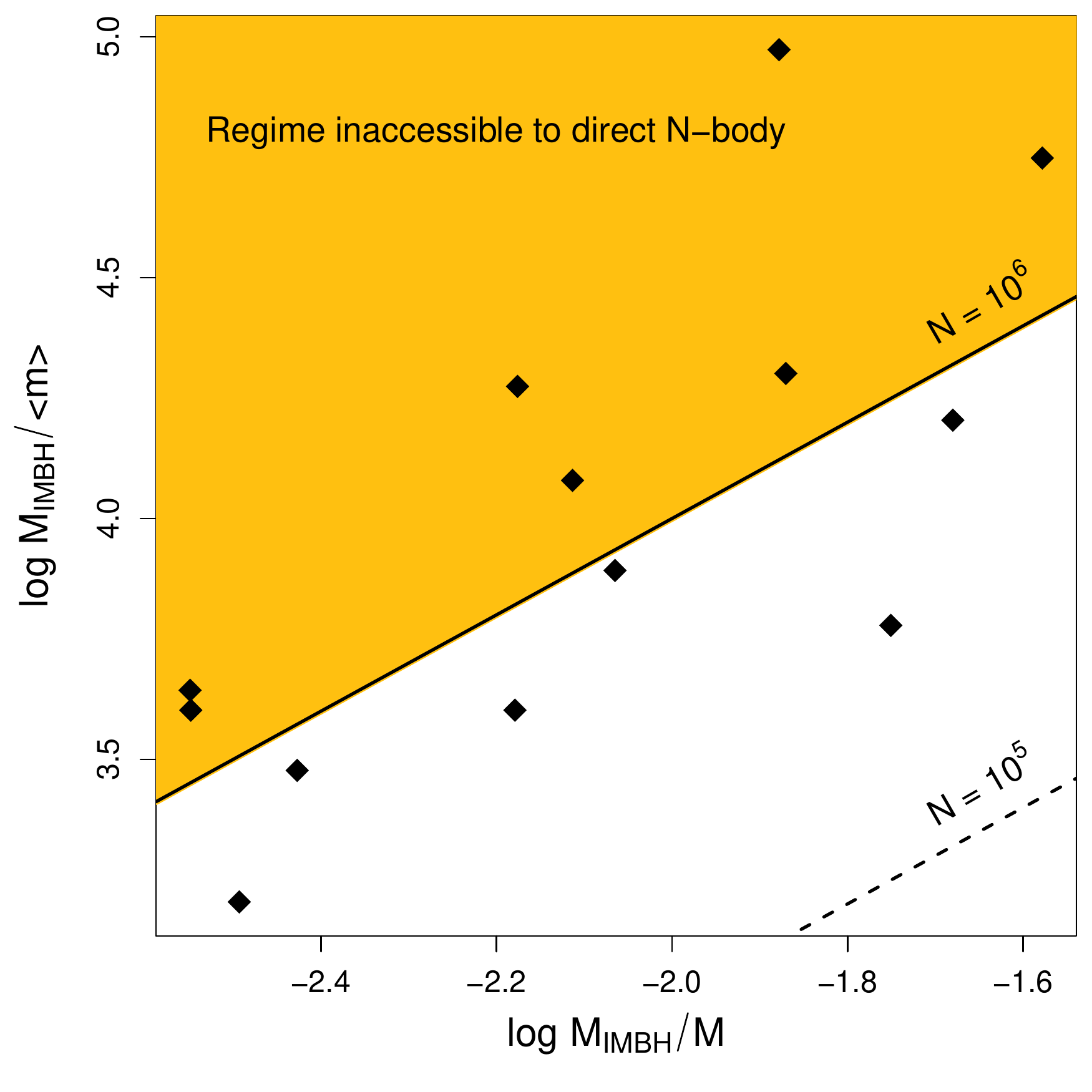}
    \caption{Observational claims of IMBH detection in the ($\mu = M_{\rm IMBH}/\langle m \rangle$, $\alpha = M_{\rm IMBH}/M$) plane, shown as black diamonds. Any technique (most notably direct N-body) which cannot simulate more than a given number of particles $N$ must trade off $\alpha$ for $\mu$ at any given $N$, being unable to access the top left region, shown shaded in orange for a typical value of $N$ for direct N-body codes currently available, $N=10^6$. As a reference $N=10^5$ is also shown as a dashed line.}
    \label{nbodyregime}
\end{figure}
%%%%%%%%%%%%%%%%%%%%%%%%%%%%%%%%%%%%%%%%%%%%%%%%%%%%%%%%
In other words, the sphere of influence is the region within which the motion of a star is heavily affected by the presence of the IMBH, while the wander radius is the typical distance at which we expect to find the IMBH from the bottom of the cluster potential well.\\
\indent Seeing how $r_{\rm inf}$ and $r_{\rm wan}$ describe two different but equally important aspects of the IMBH-host interaction and how they depend on the two ratios introduced above, we must conclude that a simulation must match both $\alpha$ and $\mu$ of a real star cluster to be considered a realistic model thereof.
However, the number of particles in a simulation is by definition
\begin{equation}
N \equiv \frac{M}{\langle m \rangle} = \frac{\mu}{\alpha}
\end{equation}
so a constraint on the number of particles $N$ that can be simulated via direct N-body translates to an inaccessible region in the ($\mu$, $\alpha$) plane.\\
\indent In Fig.~\ref{nbodyregime} we show observational claims of IMBH detections in Milky Way GCs in the ($\mu$, $\alpha$) plane. IMBH masses are based on Tab.~1 of \cite{2017IJMPD..2630021M}, which includes maximum masses in case of a negative claim. The average stellar mass in each system is assumed to be $0.5$ $M_\odot$ and the total GC mass is taken from \cite{2018MNRAS.478.1520B}. As argued above, a correct modelling of the dynamical effects of an IMBH on its host GC must match both mass ratios. Fig.~\ref{nbodyregime} shows that this is simply not possible for about half of the claims reported by \cite{2017IJMPD..2630021M} unless simulations are run with at least a million stellar particles. This is at the limit of current direct $N$-body simulation state of the art.\\
\indent An example may help clarify the meaning of Fig.~\ref{nbodyregime}. If a simulation contains $10^5$ particles, and the typical mean stellar mass for an old star cluster is $0.5$ $M_\odot$, we can take the simulation as modelling a $5 \times 10^4$ $M_\odot$ star cluster with a one-to-one star to particle ratio. If we wish to study the effect of an IMBH whose size is $0.5\%$ of the total mass of this cluster, then the IMBH must have a mass of $2.5 \times 10^2$ $M_\odot$; a quite small IMBH in comparison to the stars it interacts with.
Alternatively, we can imagine that the stellar particles in our simulation do not track actual stars one-to-one, but then any claims that direct N-body models are more realistic than other approximate methods become untenable, and it is harder to justify applying stellar evolution to each stellar particle as if it were an individual star.\\
\indent In this article we thus leverage our recently introduced MPCDSS code, which treats two- and multiple-body collisions in an approximate fashion, to simulate large-$N$ systems with a $1$-to-$1$ particle to star ratio while correctly modelling the IMBH-host interaction in the sense discussed above.\\
\indent In \cite{2022A&A...659A..19D} we observed that while the presence of a mass spectrum generally speeds up the core collapse of a given model with respect to the parent system with same total mass and $N$ but with equal mass particles, a central IMBH typically induces a shallower core collapse. It remains to determine how the velocity dispersion and density profiles are affected by the presence of a central IMBH, and more relevantly how the does the anisotropy profile evolve. Here we performed additional numerical experiments with a broader range of IMBH masses and different initial anisotropy profiles for equilibrium models with equal masses and Salpeter IMF.\\
\indent The paper is structured as follows: In Section 2 we discuss the simulation set-up, the generation of the initial conditions and we introduce the structure of MPCDSS. In Section 3 we present the results on the evolution of dynamical models of star clusters with a central IMBH. Finally, Section 4 summarizes.   
\section{Simulations}
\label{simulazij}
\subsection{Initial conditions}
In this work we have run a set of hybrid numerical simulations with our MPCDSS code discussed in DC2020, that combine a standard particle-mesh approach for the stellar potential with a multi-particle collision scheme for the collisions to a direct $N-$body code for the dynamics of the BH(s). The direct N-body treatment of the BH is a negligible overhead with respect to pure MPCDSS since the BH is only one body.\\
\indent We performed simulations with $3\times 10^3\leq N\leq 10^6$, and adopted as initial condition  the usual \cite{1911MNRAS..71..460P} profile
\begin{equation}\label{plummer}
\rho(r)=\frac{3}{4\pi}\frac{Mr_s^2}{(r_s^2+r^2)^{5/2}},
\end{equation}
with total mass $M$ and scale radius $r_s$ set to unity. Particle masses $m_i$ are either equal to $M/N$ or extracted from a \cite{1955ApJ...121..161S} power-law mass function with exponent $-2.35$ truncated such that the minimum-to-maximum-mass ratio $\mathcal{R}=m_{\rm min}/m_{\rm max}$ equals $10^{-3}$.\\
\indent In the runs without a central IMBH, we extracted the initial particles' velocities using the rejection method on the numerically recovered phase-space distribution function with  Osipkov-Merritt (hereafter OM, \citealt{1979SvAL....5...42O,1985AJ.....90.1027M}) radial anisotropy defined by 
\begin{equation}\label{OM}
f(Q)=\frac{1}{\sqrt{8}\pi^2}\int_Q^{0}\frac{{\rm d}^2\rho_a}{{\rm d}\Phi^2}\frac{{\rm d}\Phi}{\sqrt{\Phi-Q}}.
\end{equation} 
In the equation above $Q=\mathcal{E}+{J^2}/{2r_a^2}$, with $\mathcal{E}$ and $J$ the particle's energy and angular momentum per unit mass\footnote{We note that, by doing so we are assuming that the degree of anisotropy is {\it independent} of the specific particle mass. In principle, it would be also possible to generate initial conditions where different masses are associated to different degrees of radial anisotropy, e.g. see \cite{2015MNRAS.454..576G}, see also \cite{2022arXiv221206847W}.}, respectively, $\Phi$ is the gravitational potential of the model,
$r_a$ is the anisotropy radius, and $\rho_a$ the augmented density, defined by 
\begin{equation}\label{augmented}
\rho_a(r)\equiv\left(1+\frac{r^2}{r_a^2}\right)\rho(r).
\end{equation}
The anisotropy radius $r_a$ is the control parameter associated to the extent of velocity  anisotropy of the model, so that, for a given density profile the velocity-dispersion tensor is nearly isotropic inside $r_a$, and more and more radially anisotropic for increasing $r$. In other words, small values of $r_a$ correspond to more radially anisotropic systems, and thus to larger values of the anisotropy parameter $\xi$ (see e.g. \citealt{2008gady.book.....B}) defined by
\begin{equation}\label{index}
\xi\equiv\frac{2K_r}{K_t},
\end{equation}
where $K_r$ and $K_t=K_\theta+K_\phi$ are the radial and tangential components of the kinetic energy tensor that read
\begin{equation}
K_r=2\pi\int\rho(r)\sigma^2_r(r)r^2{\rm d}r,\quad K_t=2\pi\int\rho(r)\sigma^2_t(r)r^2{\rm d}r,
\end{equation} 
where, $\sigma^2_r$ and $\sigma^2_t$ are the radial and tangential phase-space averaged square velocity components, respectively.\\
\indent For the Plummer density distribution (\ref{plummer}), $f(Q)$ is given explicitly in terms of elementary functions (e.g. see  \citealt{1987MNRAS.224...13D,2017MNRAS.471.2778B}) as
\begin{equation}\label{dfplummer}
f_P(Q)=\frac{\sqrt{2}}{378\pi^2Gr_s^2\sigma_0}\left(-\frac{Q}{\sigma_0^2}\right)^{7/2}\left[1-\frac{r_s}{r_a}+\frac{63r_s^2}{4r_a^2}\left(-\frac{Q}{\sigma_0^2}\right)^{-2}\right]
\end{equation}
where $\sigma_0=\sqrt{GM/6r_s}$ is the (scalar) central velocity dispersion.\\
\indent In the simulations featuring a central IMBH, again we extract the particles positions from the density distribution (\ref{plummer}). The correspondent velocities are sampled from the standard phase-space distribution given by Eq. (\ref{dfplummer}) if their radial position $r$ is larger than the influence radius $r_{\rm inf}$, while instead for $r\leq r_{\rm inf}$ the velocities are generated sampling the isotropic distribution function $f(\mathcal{E})$ for a homogeneous and non-interacting ``atmosphere`` of density $\rho_0=3M/4\pi r_s^3$ and radius $r_{\rm inf}$ in equilibrium in $\Phi_{\rm IMBH}$, that reads
\begin{equation}
%f(\mathcal{E})=\frac{\rho_0}{2\pi^2\sqrt{2}}\left(-\mathcal{E}-\frac{GM_{\rm IMBH}}{r_{\rm inf}+\eta}\right)^{-3/2}.
f(\mathcal{E})=\frac{\rho_0}{2\pi^2\sqrt{2}}\left(-\mathcal{E}\right)^{-3/2};\quad {\rm for}\quad \mathcal{E}\leq-\frac{GM_{\rm IMBH}}{r_{\rm inf}}.
\end{equation}
For the specific choice of a Plummer density profile, the influence radius is given as function of the Plummer's scale radius $r_s$ and the IMBH mass in units of the mass ratio $\alpha$ as
\begin{equation}
r_{\rm inf}=\alpha r_s\sqrt{\frac{1}{1-\alpha^2}}.    
\end{equation}
We note that, in previous work (e.g. see \citealt{2002PhRvL..88l1103C,2002ApJ...572..371C}) the initial conditions for the stellar component have always been set up by sampling the phase-space distribution for a Plummer model without the BH and later renormalized so that the resulting systems stars+BH system is virialized. We note also that, in principle, a system with a cored density profile (such as the Plummer used here) can not have a consistent equilibrium phase-space distribution when embedded in an external potential associated to a singular density profile (such as that of the central BH, see \citealt{1996ApJ...471...68C}).\\
\indent In line with all these works, we simulate the IMBH by adding a particle, initially sitting at rest in the centre of the system with mass $10^{-4}M\leq M_{\rm IMBH}\leq 10^{-2}M$. For the range of simulation particles $3\times 10^3 \leq N \leq 10^6$, such choice corresponds to a range in mass ratios $3\leq\mu\leq 3\times10^3$. In Tab. \ref{tab_ini} we summarize the parameter of the simulations discussed in the next Sections. 
%%%%%%%%%%%%%%%%%%%%%%%%%%%%%TABLE%%%%%%%%%%%%%%%%%%%%%%%%%%%%%%%%%%%%%%%
\begin{table}
\caption{Summary of the initial conditions: After the name of each simulation (Col. 1) we report the number of simulation particles (Col. 2), the mass function (S for Salpeter, E for equal masses, Col. 3), the initial anisotropy parameter ($\xi_0$, Col. 4), the mass ratio $\alpha$  (Col. 5) and the mass ratio $\mu$ (Col. 6).} 
\begin{tabular}{llllll}
\hline
Name & $N$ & ${\rm MF}$ & $\xi_0$ & $\alpha$ & $\mu$   \\
\hline 
\texttt{s1e5xi1} & $10^5$ & S & $1.0$ & $/$ & $/$  \\
\texttt{s1e5xi2.5} & $10^5$ & S & $2.5$ & $/$ & $/$  \\
\texttt{s1e5m10xi1} & $10^5$ & S & $1.0$ & $10^{-4}$ & $10$  \\
\texttt{s1e5m30xi1} & $10^5$ & S & $1.0$ & $3\times 10^{-4}$ & $30$  \\
\texttt{s1e5m100xi1} & $10^5$ & S & $1.0$ & $10^{-3}$ & $100$  \\
\texttt{s1e5m300xi1} & $10^5$ & S & $1.0$ & $3\times 10^{-3}$ & $300$  \\
\texttt{s1e5m1e3xi1} & $10^5$ & S & $1.0$ & $10^{-2}$ & $10^{3}$  \\
\texttt{e1e5xi1} & $10^5$ & E & $1.0$ & $/$ & $/$  \\
\texttt{e1e5xi2.5} & $10^5$ & E & $2.5$ & $/$ & $/$  \\
\texttt{e1e5m10xi1} & $10^5$ & E & $1.0$ & $10^{-4}$ & $10$  \\
\texttt{e1e5m30xi1} & $10^5$ & E & $1.0$ & $3\times 10^{-4}$ & $30$  \\
\texttt{e1e5m100xi1} & $10^5$ & E & $1.0$ & $10^{-3}$ & $100$  \\
\texttt{e1e5m300xi1} & $10^5$ & E & $1.0$ & $3\times 10^{-3}$ & $300$ \\
\texttt{e1e5m1e3xi1} & $10^5$ & E & $1.0$ & $10^{-2}$ & $10^{3}$  \\
\texttt{e1e6m100xi1} & $10^6$ & E & $1.0$ & $10^{-4}$ & $100$  \\
\texttt{e1e6m300xi1} & $10^6$ & E & $1.0$ & $3\times 10^{-4}$ & $300$  \\
\texttt{e1e6m1e3xi1} & $10^6$ & E & $1.0$ & $10^{-3}$ & $10^3$  \\
\texttt{e1e6m3e3xi1} & $10^6$ & E & $1.0$ & $3\times 10^{-3}$ & $3000$ \\
\texttt{e1e6m1e4xi1} & $10^6$ & E & $1.0$ & $10^{-2}$ & $10^{4}$  \\
\texttt{s1e6xi1} & $10^6$ & S & $1.0$ & $/$ & $/$  \\
\texttt{s1e6xi1.5} & $10^6$ & S & $1.5$ & $/$ & $/$  \\
\texttt{s1e6xi2.5} & $10^6$ & S & $2.5$ & $/$ & $/$  \\
\texttt{s1e6m1e3xi1} & $10^6$ & S & $1.0$ & $10^{-3}$ & $10^{3}$  \\
\texttt{s1e6m1e3xi1.5} & $10^6$ & S & $1.5$ & $10^{-3}$ & $10^{3}$  \\
\texttt{s1e6m1e3xi2.5} & $10^6$ & S & $2.5$ & $10^{-3}$ & $10^{3}$  \\
\texttt{e1e6xi1} & $10^6$ & E & $1.0$ & $/$ & $/$  \\
\texttt{e1e6xi1.5} & $10^6$ & E & $1.5$ & $/$ & $/$  \\
\texttt{e1e6xi2.5} & $10^6$ & E & $2.5$ & $/$ & $/$  \\
\texttt{e1e6m1e3xi1} & $10^6$ & E & $1.0$ & $10^{-3}$ & $10^{3}$  \\
\texttt{e1e6m1e3xi1.5} & $10^6$ & E & $1.5$ & $10^{-3}$ & $10^{3}$  \\
\texttt{e1e6m1e3xi2.5} & $10^6$ & E & $2.5$ & $10^{-3}$ & $10^{3}$  \\
\texttt{s3e3m3xi1} & $3\times 10^3$ & S & $1.0$ & $10^{-3}$ & $3$  \\
\texttt{s1e4m10xi1} & $10^4$ & S & $1.0$ & $10^{-3}$ & $10$  \\
\texttt{s3e4m30xi1} & $3\times 10^4$ & S & $1.0$ & $10^{-3}$ & $30$  \\
\texttt{s1e5m100xi1} & $10^5$ & S & $1.0$ & $10^{-3}$ & $100$  \\
\texttt{s3e5m300xi1} & $3\times 10^5$ & S & $1.0$ & $10^{-3}$ & $300$  \\
\texttt{e3e3m3xi1} & $3\times 10^3$ & E & $1.0$ & $10^{-3}$ & $3$  \\
\texttt{e1e4m10xi1} & $10^4$ & E & $1.0$ & $10^{-3}$ & $10$  \\
\texttt{e3e4m30xi1} & $3\times 10^4$ & E & $1.0$ & $10^{-3}$ & $30$  \\
\texttt{e1e5m100xi1} & $10^5$ & E & $1.0$ & $10^{-3}$ & $100$  \\
\texttt{e3e5m300xi1} & $3\times 10^5$ & E & $1.0$ & $10^{-3}$ & $300$  \\
\texttt{e3e5m30xi1} & $3\times 10^5$ & E & $1.0$ & $10^{-4}$ & $30$  \\
\texttt{e3e5m100xi1} & $3\times 10^5$ & E & $1.0$ & $3.34\times 10^{-4}$ & $100$  \\
\texttt{e3e5m1e3xi1} & $3\times 10^5$ & E & $1.0$ & $3.34\times 10^{-3}$ & $1000$  \\
\texttt{e1e3m10xi1} & $10^3$ & E & $1.0$ & $10^{-2}$ & $10$  \\
\texttt{e1e3m3xi1} & $10^3$ & E & $1.0$ & $3\times10^{-3}$ & $3$  \\
\texttt{e1e4m100xi1} & $10^4$ & E & $1.0$ & $10^{-2}$ & $100$  \\
\texttt{e1e4m3xi1} & $10^4$ & E & $1.0$ & $3\times10^{-4}$ & $3$  \\
\texttt{e1e4m30xi1} & $10^4$ & E & $1.0$ & $3\times10^{-3}$ & $30$  \\
\texttt{e3.3e3m10xi1} & $3.34\times10^3$ & E & $1.0$ & $3\times10^{-3}$ & $10$  \\
\texttt{e3.3e4m10xi1} & $3.34\times10^4$ & E & $1.0$ & $3\times10^{-4}$ & $10$  \\
\texttt{e3.3e4m100xi1} & $3.34\times10^4$ & E & $1.0$ & $3\times10^{-3}$ & $100$  \\
\texttt{e3e2m3xi1} & $300$ & E & $1.0$ & $10^{-2}$ & $3$  \\
\texttt{e3e3m30xi1} & $3000$ & E & $1.0$ & $10^{-2}$ & $30$  \\
\texttt{e3e4m3xi1} & $3\times 10^3$ & E & $1.0$ & $10^{-4}$ & $3$  \\
\texttt{e3e4m300xi1} & $3\times 10^4$ & E & $1.0$ & $10^{-2}$ & $300$  \\
\hline
\end{tabular}
\label{tab_ini}
\end{table}
%%%%%%%%%%%%%%%%%%%%%%%%%%%%%%%%%%%%%%%%%%%%%%%%%%%%%%%%%%%%%%%%%%%%%%
%%%%%%%%%%%%%%%%%%%%%%%%%%%%%%%%%%%%%%%%%%%%%%%%%%%%%%%%%%%%%%%%%%%%%%%%%%%%%%%%%%%%%%%
\begin{figure}
\includegraphics[width = \columnwidth]{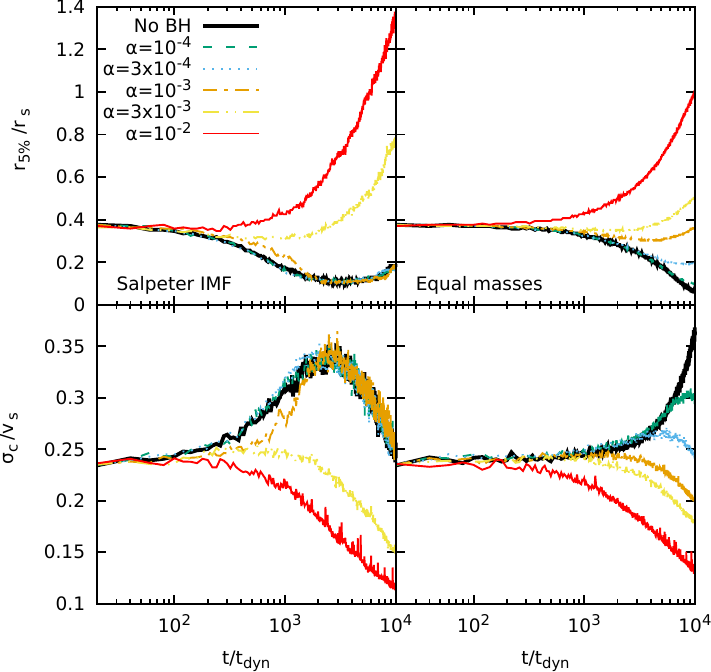}
\caption{Evolution of the Lagrangian radii $r_{5\%}$ enclosing $5\%$ of the system's mass (top panels) and central velocity dispersion $\sigma_c$ evaluated for particles inside $r_{5\%}$ (bottom panels) for models with a Salpeter mass function (left) and equal masses (right), and $\alpha =10^{-4}$, $3\times10^{-4}$, $10^{-3}$, $3\times10^{-3}$ and $10^{-2}$. In all cases the systems were initially isotropic and $N=10^5$ so that $\mu=\alpha\times 10^5$.}
\label{figsigma}
\end{figure}
%%%%%%%%%%%%%%%%%%%%%%%%%%%%%%%%%%%%%%%%%%%%%%%%%%%%%%%%%%%%%%%%%%%%%%%%%%%%%%%%%%%%%%
\subsection{Numerical scheme}
Following \cite{2021A&A...649A..24D,2022A&A...659A..19D} we evolved all sets of isolated initial conditions up to $2\times 10^4$ dynamical times $t_{\rm dyn}\equiv\sqrt{r_s^3/GM}$, so that in all cases the systems reach core collapse and are evolved further after it for at least another $10^3t_{\rm dyn}$. We employed our recent implementation of MPCDSS where the gravitational potential and force are computed by the standard particle-in-cell scheme on a fixed spherical grid
of $N_g=N_r\times N_\vartheta\times N_\varphi$ mesh points (e.g. see \citealt{1990MNRAS.242..595L}). 
%%%%%%%%%%%%%%%%%%%%%%%%%%%%%%%%%%%%%%%%%%%%%%%%%%%%%%%%%%%%%%%%%%%%%%%%%%%%%%%%%%%%%%%%%%%%%%%
%%%%%%%%%%%%%%%%%%%%%%%%%%%%%%%%%%%%%%%
\begin{figure*}
\includegraphics[width = 0.95\textwidth]{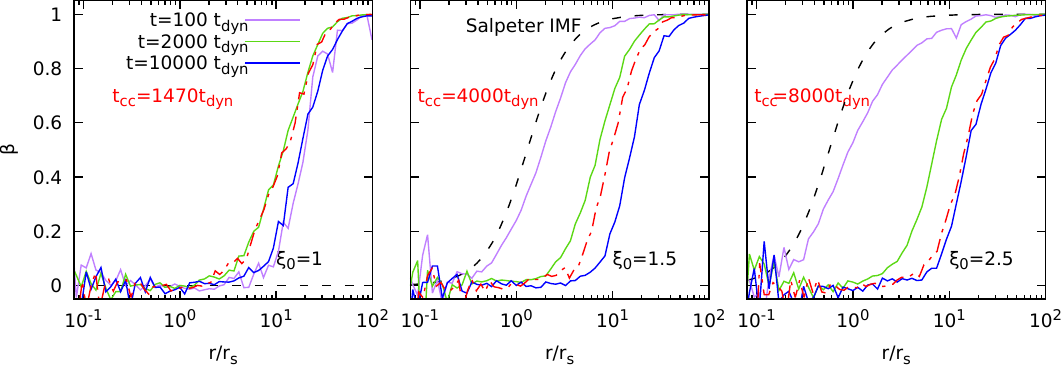}
\caption{Evolution of the Radial anisotropy profile $\beta(r)$ in Plummer models without a central IMBH, with a Salpeter IMF, $N=10^6$ and from left to right $\xi_0=1$ (isotropic), 1.5 (limit for stability) and 2.5 (critical, for consistency). The thin dashed lines mark the initial (analytical) anisotropy profile, while the dot-dashed lines mark the profile $\beta$ at the indicated time of core collapse $t_{cc}$.}
\label{figbetaevo}
\end{figure*}
%%%%%%%%%%%%%%%%%%%%%%%%%%%%%%%%%%%%%%%%%%%%%%%%%%%%%%%%%%%%%%%%%%%%%%%%%%%%%%%%%%%%%%%%%
In the simulations presented here we have used $N_r=1024$, $N_\vartheta=16$ and $N_\varphi=16$ with logarithmically spaced radial bins and averaged the potential along the azimuthal and polar coordinates in order to enforce the spherical symmetry throughout the simulation.\\
\indent The multi-particle collisions (see \citealt{2017PhRvE..95d3203D,2021A&A...649A..24D,2022A&A...659A..19D,2023IAUS..362..134D} for the details) are performed on a different mesh with $N_g=32\times 16\times 16$ extended only up to $r_{\rm cut}=100r_s$ and conditioned with a standard rejection step to the local (i.e. cell dependent) collision probability $p_i$ given by 
\begin{equation}\label{cumulative}
p_i={\rm Erf}\left(\beta\Delta t \nu_c\right),
\end{equation}
where $\Delta t$ is the simulation timestep, $\nu_c$ is the collision frequency, $\beta$ is a dimensionless constant of the order of twice the number of the simulation cells, and ${\rm Erf}(x)$ is the standard error function. 
%%%%%%%%%%%%%%%%%%%%%%%%%%%%%%%%%%%%%%%
\begin{figure*}
\includegraphics[width = 0.95\textwidth]{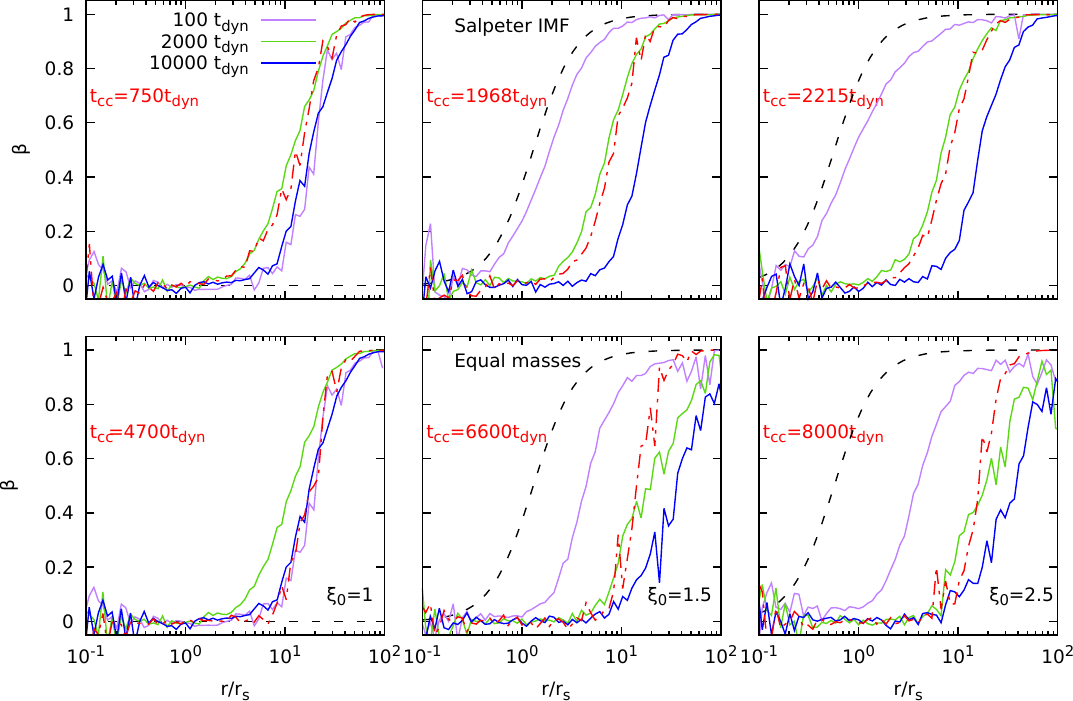}
\caption{Evolution of the radial anisotropy profile $\beta(r)$ in Plummer models with a central IMBH of mass $M_{\rm IMBH}=10^{-3}M$, with a Salpeter IMF (top panel row) or equal masses (bottom panel row), $N=10^6$ and from left to right $\xi_0=1$ (isotropic), 1.5 (limit for stability) and 2.5 (critical for consistency). The thin dashed lines mark the initial (analytical) anisotropy profile. The empirical lines become somewhat noisy to the left because of the increasingly low number of particles available for calculating $\beta(r)$ at small radii.}
\label{figbetabh}
\end{figure*}
%%%%%%%%%%%%%%%%%%%%%%%%%%%%%%%%%%%%%%%%%%%%%%%%%%%%%%%%%%%%%%%%%%%%%%%%%%%%%%%%%%%%%%%%%
In Equation (\ref{cumulative}) $\beta$ is a dimensionless constant of the order of the total number of cells in the system and 
the collision frequency is defined as usual as
\begin{equation}
\nu_c=\frac{8\pi G^2\bar{m}^2_i n_i\log\Lambda}{\sigma^3_i},
\end{equation}
where $n_i$ the local stellar number density, $\bar{m}_i$ and $\sigma_i$ the average particle mass and the putative velocity dispersion in the cell and the Coulomb logarithm $\log\Lambda$ is fixed to 10.\\
\indent In all simulations presented here we use the same normalization such that $G=M=r_s=t_{\rm dyn}=v_s=1$. Hereafter, (except where otherwise stated) all distances and velocities will be given in units of the Plummer scale radius $r_s$ and scale velocity $v_s\equiv r_s/t_{\rm dyn}$.  In our simulations, for such choice of units we adopt a constant times step $\Delta t$ and use a second order {\it leap frog} scheme to propagate the particle's equations of motion. The specific (fixed) value of the time step in units of $t_{\rm dyn}$ depends on the number of particles $N$ and their mean closest approach distance (e.g. see \citealt{2011EPJP..126...55D}) and ranges from $3\times 10^{-3}$ for $N=3\times 10^3$ to $1.25\times10^{-2}$ for $N=10^6$.\\
\indent In the runs including the central IMBH, its interaction with the stars is evaluated directly, i.e. the IMBH does not take part in the MPC step nor in the evaluation of the mean field potential. In order to keep the same rather large $\Delta t$ of the simulation, the potential exerted by the IMBH is regularized as
\begin{equation}
   \Phi_{\rm IMBH}=-\frac{GM_{\rm IMBH}}{2\epsilon}\left(3-\frac{r^2}{\epsilon^2}\right);\quad r\leq\epsilon,
\end{equation}
where we take $\epsilon=10^{-4}$ in units of $r_s$ so that for the IMBH mass-to-cluster mass ratio $10^{-3}$, the softening length is always of the order of one tenth of the influence radius of the IMBH $r_{\rm inf}$. With such choices of simulation parameters, on average, the MPC simulations on a single core are a factor $\sim 10$ faster than direct $N-$body simulations for $N$ of the order of $10^4$, and remain faster down to a factor $\sim 2$ for $N=10^6$.
%%%%%%%%%%%%%%%%%%%%%%%%%%%%%%%%%%%%%%%%%%%%%%%%%%%%%%%%%%%%%%%%%%%%%%%%%%%%%%%%%%%%%%%%
\begin{figure}
\includegraphics[width = \columnwidth]{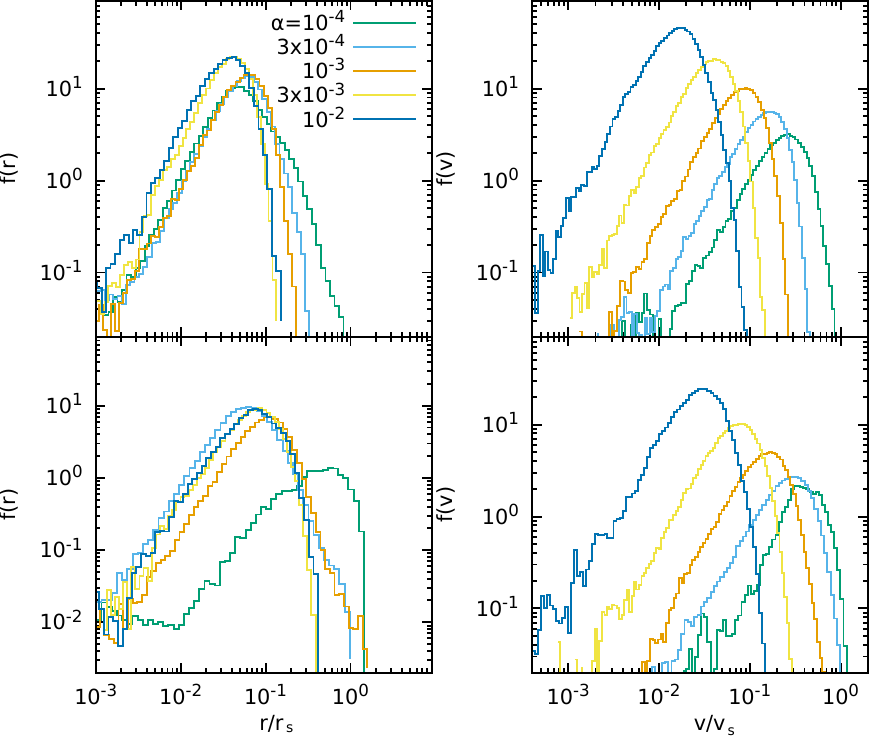}
\caption{Upper panels: Distributions of the radial coordinate (left) and velocity (right) attained by IMBHs hosted  in a Plummer model with $N=10^5$ equal mass particles with initially isotropic velocity distribution and the same combinations of $\alpha$ and $\mu$ as in Fig. \ref{figsigma}. Lower panels: Same as above but for models with Salpeter MF. \label{figdistrfrfv}}
\end{figure}
%%%%%%%%%%%%%%%%%%%%%%%%%%%%%%%%%%%%%%%%%%%%%%%%%%%%%%%%%%%%%%%%%%%%%%%%%%%%%%%%%%%%%%%%
\section{Results}
\subsection{Evolution of density and central velocity dispersion}
As an indicator of the evolution of the concentration of a given system, we followed the evolution of the Lagrangian radii containing a given fraction of the system's mass and within such radii we also computed the mean velocity dispersion. In Figure \ref{figsigma} (upper panels) we show the evolution of the Lagrangian radius $r_{5\%}$ enclosing the $5\%$ of a $N=3\times10^5$ initially isotropic (i.e. $\xi_0=1$) Plummer model with Salpeter IMF (left panels) and equal masses (right panels) and different values of the IMBH mass ratio $\alpha$. As expected, models with the mass spectrum contract on shorter time scales (at least when the specific value of the $M_{\rm IMBH}$ is low in units of $M$), with respect to their counterparts with equal mass particles. For large values of $\alpha$, $r_{5\%}$ grows rapidly without showing signs of an earlier contraction. This implies that the presence of a massive IMBH should be associated to an inflated core (see also Fig. 9 in \citealt{2022A&A...659A..19D}).\\
\indent The evolution of the average central velocity dispersion $\sigma_c$ evaluated within $r_{5\%}$ is shown in the lower panels of Fig. \ref{figsigma} (see also last column in Tab. \ref{tab_res}). We observe that for the models with a mass spectrum, $\sigma_c$ steadily decreases for the cases with $\alpha>10^{-3}$, while it grows reaching its maximum (surprisingly independent on $\alpha$) at around $t_{cc}$ and then decreases for the systems hosting a lower mass IMBH. In models with all stars having the same mass, the behaviour of $\sigma_c$ is the same for $\alpha>10^{-3}$, while it appears somewhat more complex at lower $\alpha$s. Remarkably, $\sigma_c$ reaches different maximum values before starting to decrease,  for different values of the mass ratio $\alpha$. In other words, one can conclude that the presence of a massive IMBH in a star cluster long after its core collapse, should induce a colder and larger core with respect to a star cluster in the same mass range but without a central IMBH.
\subsection{Radial anisotropy profiles}
\label{resultazij}
Before exploring the effects of a central IMBH on the orbital anisotropy of a given model, we studied the evolution of the anisotropy profiles for systems without an IMBH well beyond core collapse (and mass segregation).\\
\indent For OM systems characterized by different values of $N$ and $\xi$ we have evaluated at different times the radial anisotropy profile (see e.g. \citealt{2008gady.book.....B})
\begin{equation}\label{betar}
\beta(r)=1-\frac{\sigma_t^2(r)}{2\sigma_r^2(r)}.
\end{equation}
We find that, surprisingly, for all values of $N$ considered here between $3\times 10^3$ and $10^6$, all isotropic models ($\xi_0=1$, $\beta(r,0)=0$ everywhere) with a (Salpeter) mass spectrum have already evolved right before core collapse (typically at around ${\sim} 40 t_{\rm dyn}
$) in a ``isotropic core'' for $r<3r_s$ surrounded by an increasingly anisotropic halo of weakly bound particles kicked out during the process of mass segregation. At later times, the profile of $\beta$ remains relatively unchanged, as shown for the $N=10^6$ case in the left panel of Fig. \ref{figbetaevo} showing $\beta$ at $t=100$, 2000, and 10000$t_{\rm dyn}$ (solid lines) and at the time of core collapse $t_{cc}$ (thick  dotted-dashed line).\\
\indent The systems starting with initial conditions sampled from OM models with a larger degree of anisotropy (cfr. middle and right panels of Fig. \ref{figbetaevo}), remarkably become {\it less and less} radially anisotropic, with respect to their initial state, marked in figure by the thin dashed lines\footnote{We note that, in Osipkov-Merritt models, the radial profile of $\beta$ can be written explicitly as a function of the anisotropy radius $r_{a}$ as $\beta(r)=r^2/(r_a^2+r^2)$.}. 
We verified that such behaviour holds true even for other power-law mass spectra (not shown here) proportional to $m^{-0.6}$, $m^{-1}$ and $m^{-3}$. In these cases, the profile of $\beta$ at $t_{cc}$ for systems with low and large values of the mass function slope $\alpha$ are qualitatively very similar, for both highly anisotropic Plummer initial conditions, closer to the critical value of the anisotropy indicator, for consistency (i.e. $\xi_0=2.5$) and moderately anisotropic initial conditions (i.e. $\xi_0=1.5$); both cases showing almost isotropic ``cores'' up to $r\approx 10$.\\
\indent In practice, independently on the specific values of $\xi_0>1$, the anisotropy radius of the model increases with time as the system undergoes core collapse and re-expands. Vice versa, initially isotropic star clusters do ``anisotropize'' during core collapse and mass segregation, though their anisotropy radius also increases for $t\gg t_{cc}$. Isotropic equal masses models (not shown here), having substantially longer core collapse time scales in absence of a mass spectrum (see column 2 in Tab. \ref{tab_res}), remain substantially isotropic everywhere (i.e. with final anisotropy radii usually at about $5r_{50\%}$) while OM-anisotropic equal masses models also experience an increase in anisotropy up to core collapse as their multi-mass counterparts. We observe that, in general, the models have longer core collapse time scale at fixed mass for increasing values of the initial anisotropy parameter $\xi_0$, independently of the specific mass spectrum.
%%%%%%%%%%%%%%%%%%%%%%%%%%%%%TABLE%%%%%%%%%%%%%%%%%%%%%%%%%%%%%%%%%%%%%%%
\begin{table*}
\caption{Summary of the simulation properties: After the name of each simulation (Col. 1) we report the time of core collapse (Col. 2), the fraction of escapers at $t=10^4t_{\rm dyn}$ (Col. 3), the fraction of escapers in the last mass bin (i.e. $m>25\langle m\rangle$) at $t=10^4t_{\rm dyn}$ (Col. 4), the estimated IMBH wander radius  (Col. 5) and typical velocity (Col. 6), the core density at core collapse (Col. 7) and the central velocity dispersion at core collapse (Col. 8).} 
\begin{tabular}{llllllll}
\hline
Name & $t_{cc}/t_{\rm dyn}$ & $\%_{\rm esc}$ & $\%_{\rm esc,C}$ & $r_{\rm wan}/r_s$ & $\tilde{v}_{\rm IMBH}/v_{\rm typ}$ & $\rho_c/r_s^{-3}$ & $\sigma_c/v_{\rm typ}$   \\
\hline 
\texttt{s1e5xi1}     & $1.46\times10^3$ & $22.8\%$ & $14\%$ & $/$ & $/$ & $42.10$ & $0.234$  \\
\texttt{s1e5xi2.5}   & $1.54\times10^3$ & $24.5\%$ & $15\%$ & $/$ & $/$ & $42.80$ & $0.233$ \\
\texttt{s1e5m10xi1}  & $1.30\times10^3$ & $16\%$ & $14.7\%$ & $0.545$ & $0.311$ & $41.26$ & $0.238$ \\
\texttt{s1e5m30xi1}  & $1.79\times10^3$ & $22.8\%$ & $13.5\%$ & $0.082$ & $0.400$ & $12.61$ & $0.233$\\
\texttt{s1e5m100xi1} & $1.29\times10^3$ & $20.6\%$ & $16.1\%$ & $0.110$ & $0.167$ & $3.250$ & $0.224$  \\
\texttt{s1e5m300xi1} & $0.62\times10^3$ & $18.9\%$ & $19.5\%$ & $0.082$ & $0.075$ & $1.520$ & $0.223$ \\
\texttt{s1e5m1e3xi1} & $0.24\times10^3$ & $23.1\%$ & $35.5\%$ & $0.071$ & $0.031$ & $1.000$ & $0.222$  \\
\texttt{e1e5xi1}     & $3.24\times10^3$ & $6.1\%$ & $/$ & $/$ & $/$ & $0.860$ & $0.222$ \\
\texttt{e1e5xi2.5}   & $3.01\times10^3$ & $6.0\%$ & $/$ & $/$ & $/$ & $20.01$ & $0.250$ \\
\texttt{e1e5m10xi1}  & $8.50\times10^3$ & $8.0\%$ & $/$ & $0.075$ & $0.221$ & $38.50$ & $0.271$ \\
\texttt{e1e5m30xi1}  & $7.10\times10^3$ & $7.7\%$ & $/$ & $0.058$ & $0.166$ & $2.550$ & $0.232$ \\
\texttt{e1e5m100xi1} & $3.48\times10^3$ & $9.9\%$ & $/$ & $0.066$ & $0.097$ & $0.805$ & $0.224$  \\
\texttt{e1e5m300xi1} & $0.78\times10^3$ & $15.1\%$ & $/$& $0.055$ & $0.040$ & $0.450$ & $0.220$ \\
\texttt{e1e5m1e3xi1} & $0.10\times10^3$ & $24.9\%$ & $/$ & $0.049$ & $0.017$ & $0.350$ & $0.220$  \\
\texttt{e1e6m100xi1} & $>10^4$ & $0.6\%$ & $/$ & $0.044$ & $0.060$ & $1.000$ & $0.234$  \\
\texttt{e1e6m300xi1} & $9.51\times 10^3$ &  $1.17\%$ & $/$ & $0.038$ & $0.057$ & $0.717$ & $0.225$ \\
\texttt{e1e6m1e3xi1} & $4.50\times 10^3$ &  $3.70\%$ & $/$ & $0.041$ & $0.043$ & $0.340$ & $0.205$ \\
\texttt{e1e6m3e3xi1} & $10^3$ & $9.8\%$ & $/$ & $0.037$ & $0.027$ & $0.300$ & $0.200$ \\
\texttt{e1e6m1e4xi1} & $0.5\times 10^3$  & $10.6\%$ & $/$ & $0.016$ & $0.008$ & $0.569$ & $0.230$ \\
\texttt{s1e6xi1} & $1.47\times 10^3$ & $8.7\%$ & $6.9\%$ & $/$ & $/$ &  $20.50$  & $0.223$ \\
\texttt{s1e6xi1.5} & $4\times 10^3$  & $8.7\%$ & $7.1\%$ & $/$ & $/$ &  $21.45$  & $0.221$\\
\texttt{s1e6xi2.5} & $8\times 10^3$  & $9\%$   & $7.2\%$ & $/$ & $/$ &  $20.50$  & $0.202$\\
\texttt{s1e6m1e3xi1} & $0.75\times 10^3$ &  $7.4\%$ & $6.7\%$ & $0.072$ & $0.100$ &  $0.910$  & $0.220$  \\
\texttt{s1e6m1e3xi1.5} & $1.97\times 10^3$ &  $7.5\%$ & $6.9\%$ & $0.068$ & $0.098$ &  $0.971$  & $0.230$ \\
\texttt{s1e6m1e3xi2.5} & $2.22\times 10^3$ &  $7.6\%$ & $7.0\%$ & $0.083$ & $0.121$ &  $1.010$  & $0.210$ \\
\texttt{e1e6xi1} & $>10^4$ &  $0.6\%$ & $/$ & $/$ & $/$ &  $1.100$  & $0.265$ \\
\texttt{e1e6xi1.5}  & $>10^4$ & $0.45\%$ & $/$ & $/$ & $/$ &  $0.775$  & $0.264$ \\
\texttt{e1e6xi2.5} & $>10^4$ & $0.41\%$ & $/$ & $/$ & $/$ &  $0.550$  & $0.244$ \\
\texttt{e1e6m1e3xi1} & $4.7\times 10^3$ & $3.7\%$ & $/$ & $0.035$ & $0.042$ &  $0.360$  & $0.255$ \\
\texttt{e1e6m1e3xi1.5} & $6.6\times 10^3$ & $3.3\%$ & $/$ & $0.041$ & $0.043$ &  $0.320$  & $0.265$ \\
\texttt{e1e6m1e3xi2.5} & $8.0\times 10^3$ & $3\%$ & $/$ & $0.049$ & $0.045$ &  $0.275$  & $0.229$ \\
\texttt{s3e3m3xi1}     & $8.0\times 10^3$ & $14.9\%$  & $7\%$ & $1.680$ & $0.200$ & $5.000$ & $0.260$ \\
\texttt{s1e4m10xi1}  & $3.1\times 10^3$ & $18.7\%$ & $9.7\%$  & $0.675$ & $0.552$ & $8.000$ & $0.250$ \\
\texttt{s3e4m30xi1}  & $1.4\times 10^3$ & $27.3\%$ & $11\%$ & $0.114$ & $0.291$ & $9.831$ & $0.238$ \\
\texttt{s1e5m100xi1} & $0.8\times10^3$  & $30.7\%$ & $11.2\%$ & $0.311$ & $0.092$ & $17.61$ & $0.232$ \\
\texttt{s3e5m300xi1} & $0.7\times10^3$  & $33.5\%$ & $11.3\%$ & $0.131$ & $0.156$ & $30.10$ & $0.224$   \\
\texttt{e3e3m3xi1}   & $4.8\times10^3$  & $6.97\%$ & $/$ & $0.290$ & $0.344$ & $0.800$ & $0.250$ \\
\texttt{e1e4m10xi1}  & $4.6\times10^3$  & $12.5\%$ & $/$ & $0.191$ & $0.189$ & $1.000$ & $0.260$\\
\texttt{e3e4m30xi1}  & $6.9\times10^3$  & $13.1\%$ & $/$ & $0.141$ & $0.123$ & $21.75$ & $0.300$\\
\texttt{e1e5m100xi1} & $4.1\times10^3$  & $15.8\%$ & $/$ & $0.075$ & $0.081$ & $0.750$ & $0.245$\\
\texttt{e3e5m300xi1} & $5.1\times10^3$  & $11.1\%$ & $/$ & $0.045$ & $0.064$ & $0.450$ & $0.230$\\
\texttt{e3e5m30xi1}  & $1.6\times10^4$  & $10.67\%$ & $/$ & $0.059$ & $0.371$ & $9.500$ & $0.222$ \\
\texttt{e3e5m100xi1} & $10^4$ & $3.37\%$ & $/$ & $0.057$ & $0.231$  & $1.200$ & $0.224$\\
\texttt{e3e5m1e3xi1} & $0.4\times 10^3$ & $13.34\%$ & $/$ & $0.041$ & $0.058$ & $0.317$ & $0.225$\\
\texttt{e1e3m10xi1} & $10^3$ & $6.3\%$ & $/$ & $0.322$ & $0.232$ & $1.900$ & $0.162$\\
\texttt{e1e3m3xi1} & $9.4\times10^3$ & $1.8\%$ & $/$ & $0.218$ & $0.328$ & $2.000$ & $0.220$\\
\texttt{e1e4m100xi1} & $10^3$ & $12\%$ & $/$ & $0.170$ & $0.091$ & $0.200$ & $0.158$\\
\texttt{e1e4m3xi1} & $>10^4$ & $7.32\%$ & $/$ & $0.392$ & $0.460$ & $20.00$ & $0.265$\\
\texttt{e1e4m30xi1} & $>10^4$ & $7\%$ & $/$ & $0.229$ & $0.235$ & $18.00$ & $0.262$\\
\texttt{e3.3e3m10xi1} & $9\times10^3$ & $3.96\%$ & $/$ & $0.259$ & $0.243$ & $1.000$ & $0.225$\\
\texttt{e3.3e4m10xi1} & $7\times10^3$ & $9.68\%$ & $/$ & $0.100$ & $0.281$ & $3.000$ & $0.215$\\
\texttt{e3.3e4m100xi1} & $1.41\times10^3$ & $14.8\%$ & $/$ & $0.132$ & $0.082$ & $0.150$ & $0.141$\\
\texttt{e3e2m3xi1} & $7.88\times10^3$ & $6.67\%$ & $/$ & $0.467$ & $0.406$ & $4.000$ & $0.202$\\
\texttt{e3e3m30xi1} & $>10^4$ & $6.43\%$ & $/$ & $0.256$ & $0.146$ & $0.400$ & $0.180$\\
\texttt{e3e4m3xi1} & $9\times10^3$ & $10.3\%$ & $/$ & $0.217$ & $0.265$ & $61.00$ & $0.292$\\
\texttt{e3e4m300xi1} & $>10^4$ & $22.37\%$ & $/$ & $0.113$ & $0.039$ & $0.100$ & $0.106$\\
\hline
\end{tabular}
\label{tab_res}
\end{table*}
%%%%%%%%%%%%%%%%%%%%%%%%%%%%%%%%%%%%%%%%%%%%%%%%%%%%%%%%%%%%%%%%%%%%%%
%%%%%%%%%%%%%%%%%%%%%%%%%%%%%%%%%%%%%%%%%%%%%%%%%%%
\begin{figure}
    \centering
    \includegraphics[width=\columnwidth]{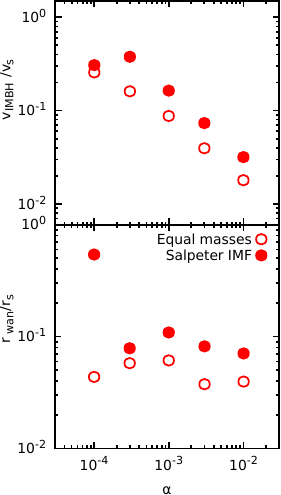}
    \caption{For single mass models (empty circles) and models with Salpeter mass function (filled circles) the most probable velocity and the putative wander radius of the IMBH, respectively, are shown as a function of the mass ratio $\alpha$. In all cases $N=10^{5}$.}
    \label{equivssalp}
\end{figure}
%%%%%%%%%%%%%%%%%%%%%%%%%%%%%%%%%%%%%%%%%%%%%%%%%%%%%%%%
Adding a central black hole, for all initial conditions discusses here, has the effect of systematically reducing $t_{cc}$ of a factor between 2 and 3.5. As observed for models without a central BH, isotropic initial conditions tend to evolve towards more anisotropic states also for the cases with the BH (see left panels in Fig. \ref{figbetabh}) with or without a mass spectrum. In the latter, the strong kicks exerted by the BH on eccentric orbits play the role of mass segregation in populating the outer radii of low angular momentum stars. OM models with a central BH again evolve towards less anisotropic states with equal mass systems with significantly flatter $\beta$ profiles at late times. Of course, the interplay between the evolution of the anisotropy profiles and that of the IMBH should be, in principle, studied in models starting from initial conditions where the proto-cluster is far from being virialized with or without a significantly massive seed for the IMBH, such as those produced in \cite{2022MNRAS.510.2097T}.  
\subsection{Wander radius}
\begin{figure}
    \centering
    \begin{tabular}{c}
    \includegraphics[width=\columnwidth]{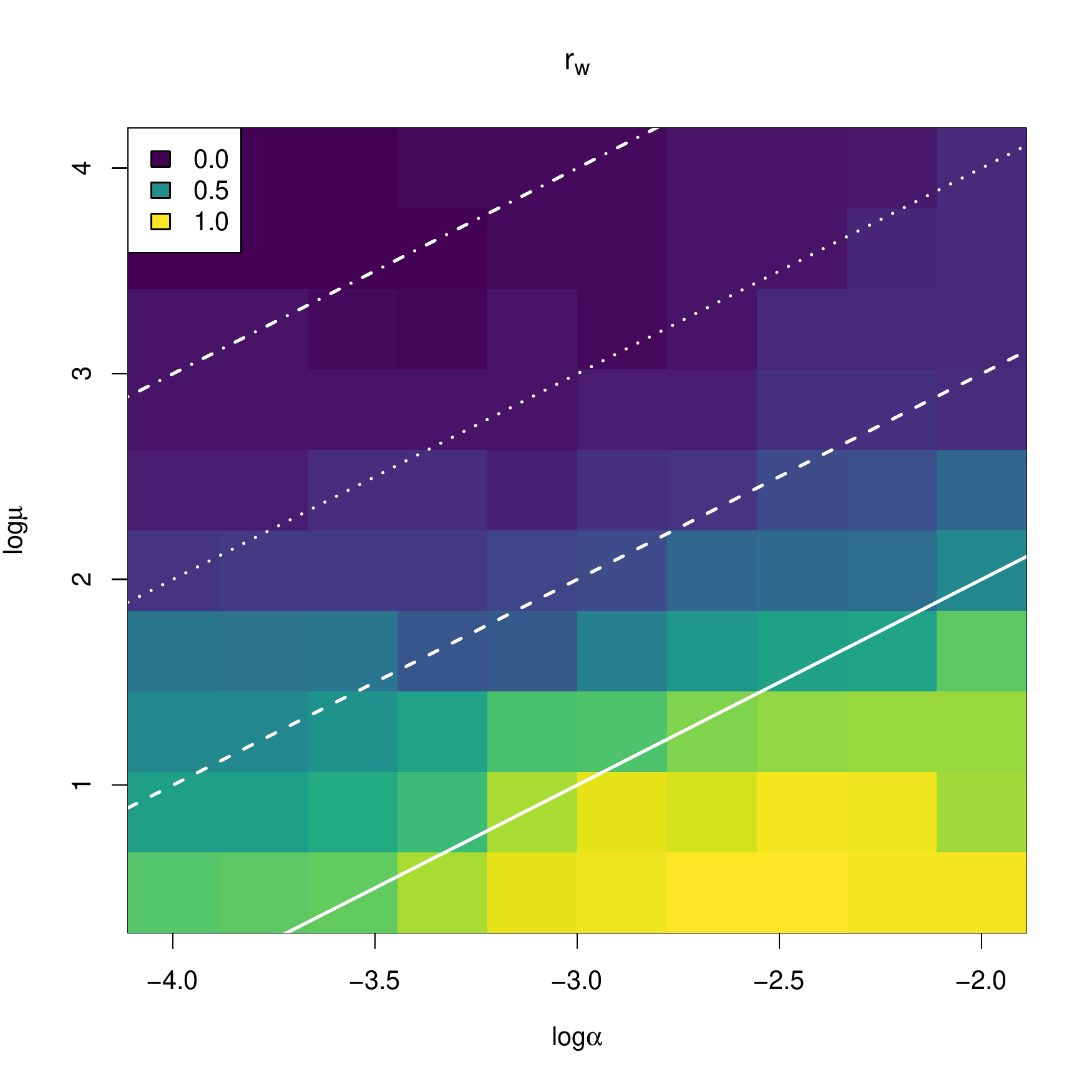}\\ \includegraphics[width=\columnwidth]{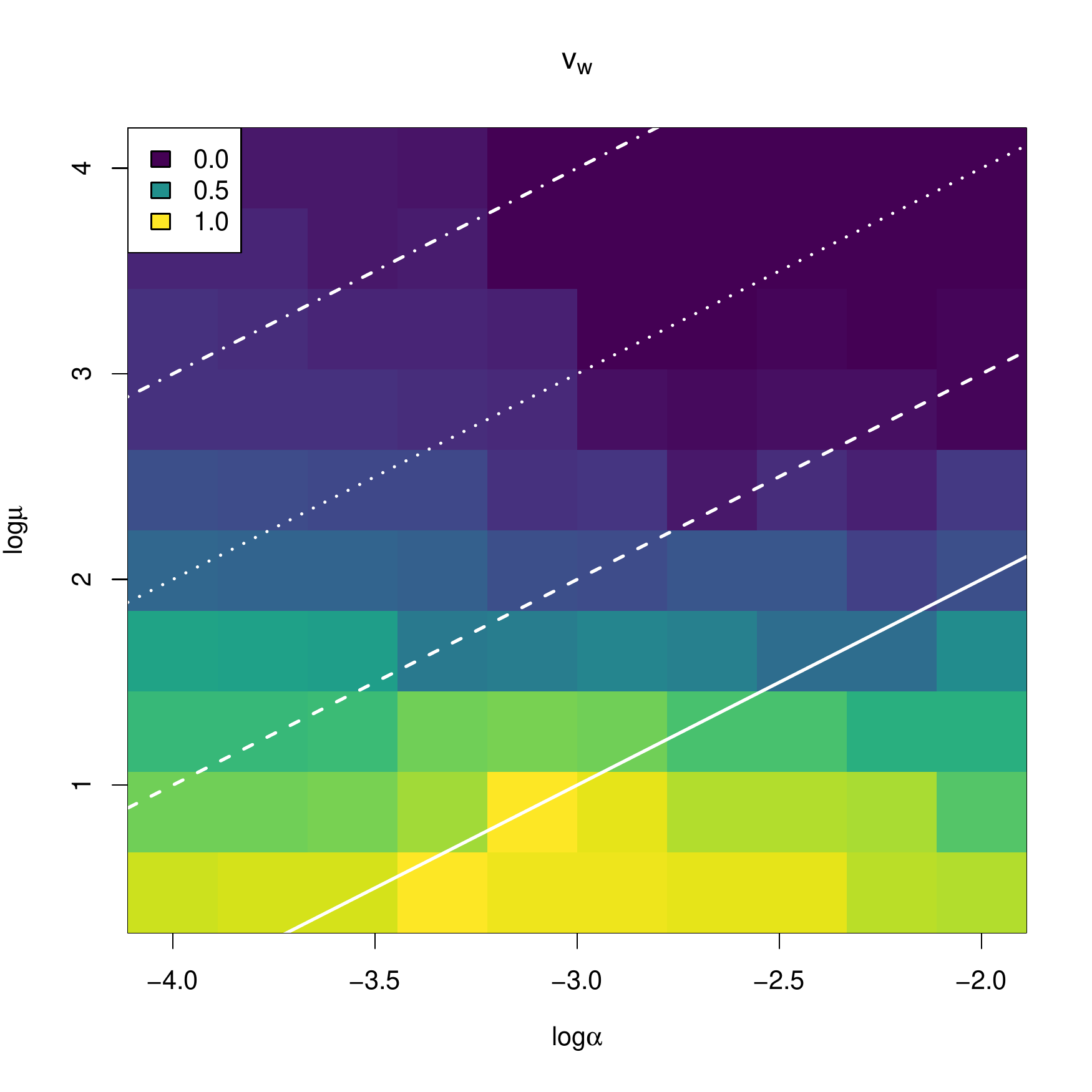}\\
    \end{tabular}
    \caption{Wander radius (scaled to $0$-$1$ over the simulation sample) as a function of $\mu$ and $\alpha$ (top panel) and wander velocity as a function of $\mu$ and $\alpha$ (bottom panel). The diagonal lines correspond to a constant number of particles: $10^4$ (solid), $10^5$ (dashed), $10^6$ (dotted), $10^7$ (dot-dashed).}
    \label{wander}
\end{figure}
%%%%%%%%%%%%%%%%%%%%%%%%%%%%%%%%%%%%%%%%%%%%%%%%%%%%%%%%%%%%%%%%%%%%%%%%%%%%%%%%%%%%%%%
\begin{figure*}
\includegraphics[width = \textwidth]{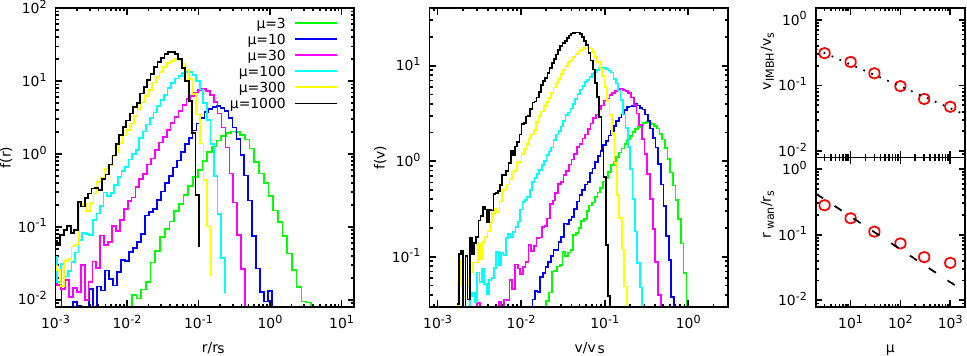}
\caption{Distributions of the radial coordinate (left panel) and velocity (middle panel) of IMBHs embedded in a Plummer model with $3\times 10^3 \leq N\leq 10^6$ equal mass particles and initially isotropic velocity distribution. The top right and top left panels show as function of the mass ratio $\mu$ the peaks of the velocity and radial distributions, respectively. $\tilde{v}_{\rm IMBH}$ shows a markedly $\mu^{-1/3}$ trend (dotted line), while $r_{\rm wan}$ has a trend compatible with $\mu^{-1/2}$ for $\mu<3\times10^2$ (dash ed line). In all cases $\alpha=10^{-3}$.}
\label{frmu}
\end{figure*}
\begin{figure}
    \centering
    \includegraphics[width=\columnwidth]{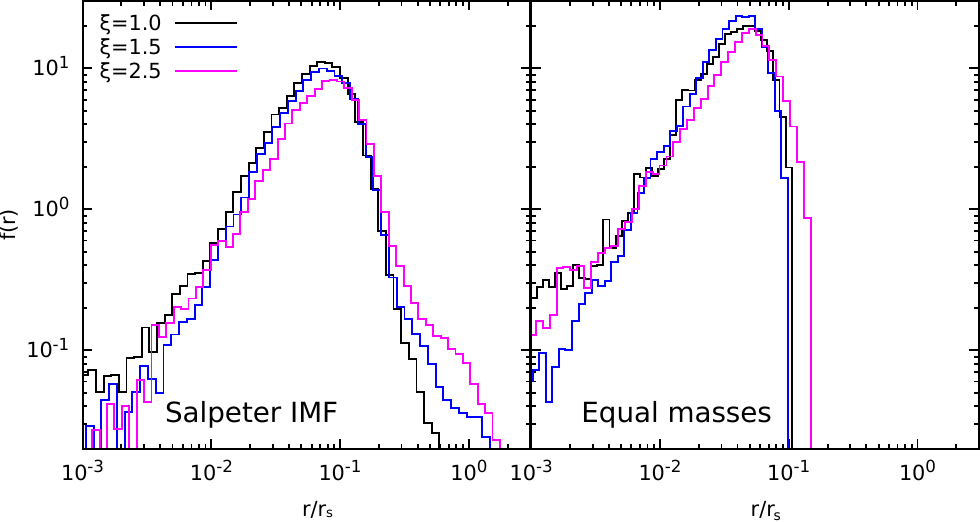}
    \caption{Distribution of the IMBH radial coordinate for Plummer models with $N=10^6$, $M_{\rm IMBH}=10^{-3}$ and $\xi=1,$ 1.5 and 2.5, for model with Salpeter mass function (left panel) and equal masses (right panel).}
    \label{anisotropic}
\end{figure}

%%%%%%%%%%%%%%%%%%%%%%%%%%%%%%%%%%%%%%%%%%%%%%%%%%%%%%%%%%%%%%%%%%%%%%%%%%%%%%%%%%%%%%%
For the IMBH hosted in the models discussed in the previous Sections, we have evaluated the probability density functions (PDF) of the radial position with respect to the geometric centre of the stellar distribution $f(r)$, and velocity $f(v)$. In Figs. (\ref{figdistrfrfv}) we show said distributions for initially isotropic models  with $N=10^6$ and $10^{-4}\leq\alpha\leq10^{-2}$ and either single-mass or Salpeter mass spectra, respectively.\\
\indent As an indicator of the extension of the BH wander radius $r_{\rm wan}$ we extract the radius corresponding to the peak of $f(r)$.\\
\indent We observe that (see Fig. \ref{equivssalp}), for fixed $N$ while the peak of the velocity distribution $f(v)$ moves at lower velocities for increasing $\alpha$ (for both equal mass and Salpeter systems), the peak of the radial position distribution $f(r)$ (i.e. the putative wander radius) is somewhat independent on $\alpha$ for equal mass models, while it moves to smaller values of $r$ for increasing $\alpha$ in systems with a mass spectrum.\\
\indent When fixing $\alpha$ while decreasing $\mu$ (i.e. the we change $N$ so that the BH mass to mean stellar mass $\langle m\rangle$ varies), both $f(r)$ and  $f(v)$ become broader and peak at larger $r$ and $v$, respectively (see Fig. \ref{frmu}, main panels). Remarkably, we recover (at least for $\mu\lesssim 200$) the predicted $r_{\rm wan}\propto \mu^{-1/2}$ trend, while we observe a clear $v_{\rm IMBH}\propto \mu^{-1/3}$ behaviour for the typical velocity of the IMBH (Dashed and dotted lines in the right panel of Fig. \ref{frmu}, respectively). Not surprisingly, for fixed $\mu$ we observe larger values of the wander radius $r_{\rm wan}$ in systems with smaller $\alpha$, as in those cases the larger cluster mass forms a deeper potential well. This is exemplified in Fig. \ref{wander} where $r_{\rm wan}$ (top panel) and $v_{\rm IMBH}$ (bottom panel) are plotted are colour coded against $\alpha$ and $\mu$ for the equal $m$ cases.\\
\indent We observe that, in general, for mass ratios $\mu$ larger than 10, all such trends are weakly affected by the mass spectrum or the specific anisotropy profile of the model at hand. However, we notice that for increasing initial values of $\xi$, the distribution of the radial coordinate of the IMBH shows systematically fatter tails, corresponding to a decreasing (negative) kurtosis $\kappa$. As an example, in Fig. \ref{anisotropic} we show $f(r)$ for $\mu=10^3$, $\alpha=10^{-3}$, $\xi_0=1,$ 1.5 and 2.5; and Salpeter (left panel) and equal mass (right panel) models. This implies that, IMBHs in models with markedly anisotropic initial conditions might have a non negligible probability of being displaced of a few scale radii from the geometric centre of the star cluster. We note that, \cite{2002PhRvL..88l1103C,2002ApJ...572..371C} by means of direct $N-$body simulations and Fokker-Planck calculations in a static cluster potential $\Phi$ estimated a limit $r_{\rm wan}$ of the order of $0.1r_s$, where $r_s$ is some scale length roughly equal to the half-mass radius of the model at hand. Such value is typically assumed as the radial distance\footnote{Such radius of about 0.1$r_s$ is also consistent with the typical core-stalling radius where dynamical friction and dynamical buoyancy compensate each other (e.g. see \cite{2021ApJ...912...43B,2022ApJ...926..215B})} within which to look for IMBH candidates in GCs in many observational studies. It is important to note that, on one hand such $N-$body runs had the natural limits of the relatively small number of (equal mass) particles $N$ and their overall large computational cost. On the other hand, the Fokker-Planck models used, somewhat arbitrarily, a Gaussian force fluctuation distribution $f(\delta F)$. In both cases therefore, the rare but strong encounters where systematically neglected.\\
\indent \cite{2020IAUS..351...93D} studied the dynamics of massive BH in galactic cores using a model based on the integration of stochastic (i.e. Langevin) equations (see also \citealt{2020A&A...640A..79P}) of the form
\begin{equation}
\ddot{\mathbf{r}}_{\rm BH}=-\nabla\Phi-\eta\mathbf{v}_{\rm BH}+\delta F,
\end{equation}
where $\eta$ is the Chandrasekhar dynamical friction coefficient and $\delta F$ a fluctuating force (per unit mass). They showed that the position distribution of the BH extracted from short time $N-$body simulations is qualitatively ``intermediate'' between those obtained in longer Langevin simulations with force fluctuations sampled from a Gaussian and a \cite{1919AnP...363..577H} distribution for $f(\delta F)$ (see their Fig. 1). The latter being the correct force fluctuation distribution in a system of particles interacting with $1/r^2$ force law (\citealt{1942ApJ....95..489C,1943ApJ....97....1C}).\\
\indent We stress that fact that in the MPC simulations discussed in the present work the interactions between the IMBH and the stars are evaluated with a direct sum scheme (as in \citealt{2002PhRvL..88l1103C}), but the evolution time, being of the order of several thousands of crossing times, is much larger, thus allowing for strong encounters (typically corresponding to strong force fluctuations described by the heavy $F^{-5/2}$ tails of the Holtsmark distribution) to have a non negligible role in the dynamics of the IMBH.   
%%%%%%%%%%%%%%%%%%%%%%%%%%%%%%%%%%%%%%%%%%%%%%%%%%%
%%%%%%%%%%%%%%%%%%%%%%%%%%%%%%%%%%%%%%%%%%%%%%%%%%%%%%%%
\subsection{Escapers and compact objects retention fraction}
In \cite{2021A&A...649A..24D} we have compared the time dependent fraction of escapers (i.e. particles reaching with positive total energy a truncation radius fixed at $\sim 20r_{50\%}$) in MPC and direct simulations with isotropic initial conditions with $N$ of order $10^4$, finding a rather good agreement for several choices of the mass spectrum. Here we evaluate the fraction of escapers for a broader range of $N$ and different choices of $\xi_0$ (cfr. column 3 in Tab. \ref{tab_res}).\\
\indent In general, over a time span of $10^4t_{\rm dyn}$, models with equal masses tend to have a lower fraction of escapers than those with same $N$ and $\xi_0$ with a mass spectrum, this is ascribed to the mass segregation process that pushes heavier stars to the inner regions of the cluster lowering their potential energy, at the expense of lighter stars pushed outside with increasing kinetic energies. This is also observable in \cite{2021A&A...649A..24D} (cfr. Fig. 6 therein) where models characterized by heavier tailed mass functions (i.e. larger fractions of heavy particles at fixed $\langle m\rangle$) show a steeper time increase of the escapers fraction.\\
%%%%%%%%%%%%%%%%%%%%%%%%%%%%%%%%%%%%%%%%%%%%%%%%%%%
\begin{figure}
    \centering
    \includegraphics[width=\columnwidth]{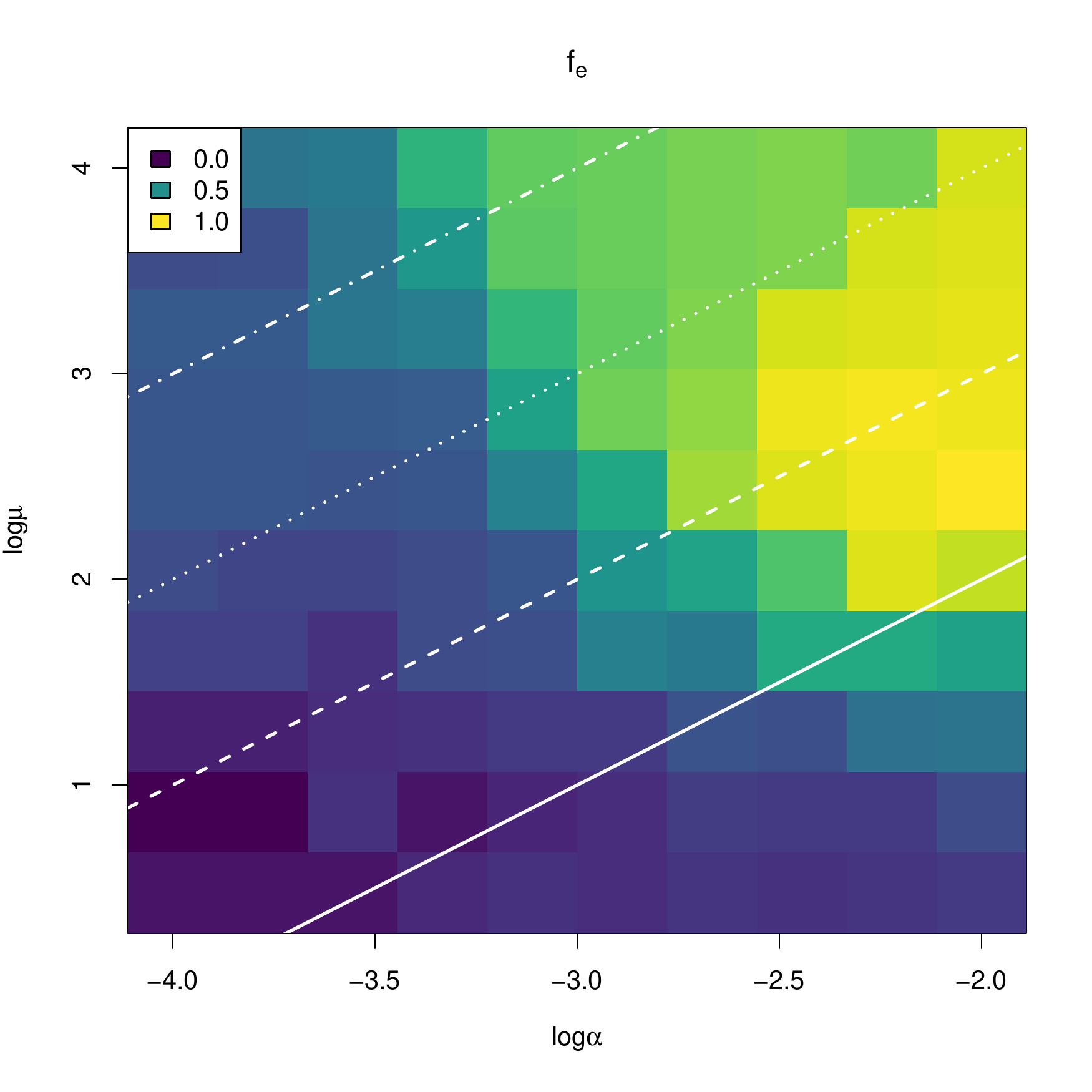}
    \caption{Fraction of escapers as a function of $\mu$ and $\alpha$ (top panel) as a function of $\mu$ and $\alpha$ (bottom panel). The diagonal lines correspond to a constant number of particles: $10^4$ (solid), $10^5$ (dashed), $10^6$ (dotted), $10^7$ (dot-dashed).}
    \label{fesc}
\end{figure}
%%%%%%%%%%%%%%%%%%%%%%%%%%%%%%%%%%%%%%%%%%%%%%%%%%%%%%%%
\indent The presence of an IMBH, even if associated with a shallower core collapse, typically enhances particle evaporation via direct collisions with larger escapers fractions for increasing values of the mass ratio $\mu$. For the models with or without an IMBH the initial anisotropy profile has little influence on the fraction of escapers and no apparent trend is evident, as the latter might depend on $\mu$, $\xi$ and $\alpha$ simultaneously (see Fig. \ref{fesc} below).\\
\indent For the models with Salpeter mass function we have also evaluated the mass-dependent escaper fraction, dividing the mass spectrum of the system in 50 logarithmically spaced mass bins. In column 4 of Tab. \ref{tab_res} we give the fraction of escapers in the largest bin (corresponding roughly to $m> 25\langle m\rangle$). For star clusters with a mean stellar mass of about $0.5M_\odot$ these would correspond to $m>10M_{\odot}$, likely encompassing collapsed objects. Not surprisingly, bigger values of $\mu$ are associated to increasing fractions of heavy escapers. In the worst case, up to the $35\%$ of particles in the largest mass bin are ejected before $10^4t_{\rm dyn}$, that corresponds to a compact object retention fraction of about $65\%$. Again, the initial anisotropy profile does not have a significant effect on the retention fraction for fixed values of $\mu$ or $\alpha$.
\section{Discussion and conclusions}
We investigated with multi-particle collision simulations with MPCDSS the dynamics of IMBHs in star clusters under different characteristics of the host (mass spectrum, orbital anisotropy) and the IMBH itself (mass ratio to the typical star and to the total host mass). Thanks to the linear complexity of MPCDSS with the number of particles, we had the opportunity to explore a wider range in these mass ratios as discussed in the introduction.\\
\indent We confirmed our preliminary results of \cite{2022A&A...659A..19D} that the presence of a central black hole of mass about $10^{-3}$ in units of the total cluster mass induces a more a faster but shallower core collapse. This remains true for other values of the mass rations $\mu$ and $\alpha$ defined in Sect. 1.\\
\indent In practice, clusters hosting a central IMBH would be observed as ``dynamically older'' than their counterparts with no BH and with a more diffuse and colder core. Moreover for fixed mean stellar mass, all systems with the central BH have a significantly larger fraction of escapers (and a smaller retention fraction of heavier stars) than those with no BH. We have explored the effect of Osipkov-Merritt initial anisotropy profiles finding that long after the core collapse time has been reached, independently on the initial value of $\xi$ and the presence or not of an IMBH, the clusters show a anisotropy profile with $\beta$ between 2 and 10 initial scale radii $r_s$, or of about 5 final half mass radii $r_{50\%}$.\\
\indent We have evaluated the PDF of the radial displacement of the IMBH $f(r)$ (i.e. the distribution of its radial distance from the geometric centre of the star cluster) and defined as IMBH wander radius its absolute maximum. If on one hand we recover the $(M_{\rm IMBH}/\langle m\rangle)^{-1/2}$ trend (independently of the cluster mass and anisotropy profile), on the other we observe that such radius is seemingly less dependent on the $\alpha$ ratio, being typically of the order of $10^{-1}r_{50\%}$. A result whose importance cannot be overstated is that the distribution of the distance as well as the sky-projected distance from the centre attained by the IMBH in our simulations becomes distinctly leptokurtic for increasing values of the systems' initial anisotropy. This corresponds to the presence of heavy tails, with the associated risk of underestimating the probability of low probability events. A possible astrophysical consequence could be the unduly exclusion of potential IMBH candidates when they happen to be too far away from the host systems' centre based on our Gaussian/Brownian expectations.
\cite{2018ApJ...862...16T} for instance exclude several radio sources from their analysis even though they are relatively near to the host star cluster centre, because they are further out than the estimated Brownian radius of an IMBH of the relevant mass.  
%%%%%%%%%%%%%%%%%%%%%%%%%%%%%%%%%%%%%%%%%%%%%%%%%%%%%%%%%%%%%5
\begin{acknowledgements}
This material is based upon work supported by the ``Fondazione Cassa di Risparmio di Firenze'' under the project {\it HIPERCRHEL} for the use of high performance computing resources at the university of Firenze. P.F.D.C. is supported by the MIUR-PRIN2017 project \textit{Coarse-grained description for non-equilibrium systems and transport phenomena
(CO-NEST)} n.201798CZL. L.B. is financed by the ``Fondazione Cassa di Risparmio di Firenze'' under the project {\it THE SWITCH}. M.~P. acknowledges financial support from the European Union’s Horizon 2020 research and innovation programme under the Marie Sklodowska-Curie grant agreement No. $896248$. We warmly acknowledge Evangelia Tremou for reading a draft of this manuscript.
\end{acknowledgements}
%%%%%%%%%%%%%%%%%%%%%%%%%%%%%%%%%%%%%%%%%%%%%%%%%%%%%%%%%%55
   \bibliographystyle{aa} % style aa.bst
   \bibliography{manuscript} % your references Yourfile.bib

\begin{thebibliography}{71}
\expandafter\ifx\csname natexlab\endcsname\relax\def\natexlab#1{#1}\fi

\bibitem[{{Aros} {et~al.}(2020){Aros}, {Sippel}, {Mastrobuono-Battisti},
  {Askar}, {Bianchini}, \& {van de Ven}}]{2020MNRAS.499.4646A}
{Aros}, F.~I., {Sippel}, A.~C., {Mastrobuono-Battisti}, A., {et~al.} 2020,
  \mnras, 499, 4646

\bibitem[{{Bahcall} \& {Wolf}(1976)}]{1976ApJ...209..214B}
{Bahcall}, J.~N. \& {Wolf}, R.~A. 1976, \apj, 209, 214

\bibitem[{{Ballone} {et~al.}(2018){Ballone}, {Mapelli}, \&
  {Pasquato}}]{2018MNRAS.480.4684B}
{Ballone}, A., {Mapelli}, M., \& {Pasquato}, M. 2018, \mnras, 480, 4684

\bibitem[{{Banik} \& {van den Bosch}(2021)}]{2021ApJ...912...43B}
{Banik}, U. \& {van den Bosch}, F.~C. 2021, \apj, 912, 43

\bibitem[{{Banik} \& {van den Bosch}(2022)}]{2022ApJ...926..215B}
{Banik}, U. \& {van den Bosch}, F.~C. 2022, \apj, 926, 215

\bibitem[{{Baumgardt}(2001)}]{2001MNRAS.325.1323B}
{Baumgardt}, H. 2001, \mnras, 325, 1323

\bibitem[{{Baumgardt} {et~al.}(2008){Baumgardt}, {De Marchi}, \&
  {Kroupa}}]{2008ApJ...685..247B}
{Baumgardt}, H., {De Marchi}, G., \& {Kroupa}, P. 2008, \apj, 685, 247

\bibitem[{{Baumgardt} \& {Hilker}(2018)}]{2018MNRAS.478.1520B}
{Baumgardt}, H. \& {Hilker}, M. 2018, \mnras, 478, 1520

\bibitem[{{Binney} \& {Tremaine}(2008)}]{2008gady.book.....B}
{Binney}, J. \& {Tremaine}, S. 2008, {Galactic Dynamics: Second Edition}
  (Princeton University Press)

\bibitem[{{Bortolas} {et~al.}(2016){Bortolas}, {Gualandris}, {Dotti}, {Spera},
  \& {Mapelli}}]{2016MNRAS.461.1023B}
{Bortolas}, E., {Gualandris}, A., {Dotti}, M., {Spera}, M., \& {Mapelli}, M.
  2016, \mnras, 461, 1023

\bibitem[{{Breen} {et~al.}(2017){Breen}, {Varri}, \&
  {Heggie}}]{2017MNRAS.471.2778B}
{Breen}, P.~G., {Varri}, A.~L., \& {Heggie}, D.~C. 2017, \mnras, 471, 2778

\bibitem[{{Brockamp} {et~al.}(2011){Brockamp}, {Baumgardt}, \&
  {Kroupa}}]{2011MNRAS.418.1308B}
{Brockamp}, M., {Baumgardt}, H., \& {Kroupa}, P. 2011, \mnras, 418, 1308

\bibitem[{{Bustamante-Rosell} {et~al.}(2021){Bustamante-Rosell}, {Noyola},
  {Gebhardt}, {Fabricius}, {Mazzalay}, {Thomas}, \&
  {Zeimann}}]{2021ApJ...921..107B}
{Bustamante-Rosell}, M.~J., {Noyola}, E., {Gebhardt}, K., {et~al.} 2021, \apj,
  921, 107

\bibitem[{{Carr} {et~al.}(1984){Carr}, {Bond}, \&
  {Arnett}}]{1984ApJ...277..445C}
{Carr}, B.~J., {Bond}, J.~R., \& {Arnett}, W.~D. 1984, \apj, 277, 445

\bibitem[{{Chandrasekhar} \& {von Neumann}(1942)}]{1942ApJ....95..489C}
{Chandrasekhar}, S. \& {von Neumann}, J. 1942, \apj, 95, 489

\bibitem[{{Chandrasekhar} \& {von Neumann}(1943)}]{1943ApJ....97....1C}
{Chandrasekhar}, S. \& {von Neumann}, J. 1943, \apj, 97, 1

\bibitem[{{Chatterjee} {et~al.}(2002{\natexlab{a}}){Chatterjee}, {Hernquist},
  \& {Loeb}}]{2002PhRvL..88l1103C}
{Chatterjee}, P., {Hernquist}, L., \& {Loeb}, A. 2002{\natexlab{a}}, \prl, 88,
  121103

\bibitem[{{Chatterjee} {et~al.}(2002{\natexlab{b}}){Chatterjee}, {Hernquist},
  \& {Loeb}}]{2002ApJ...572..371C}
{Chatterjee}, P., {Hernquist}, L., \& {Loeb}, A. 2002{\natexlab{b}}, \apj, 572,
  371

\bibitem[{{Ciotti}(1996)}]{1996ApJ...471...68C}
{Ciotti}, L. 1996, \apj, 471, 68

\bibitem[{{Dehnen} \& {Read}(2011)}]{2011EPJP..126...55D}
{Dehnen}, W. \& {Read}, J.~I. 2011, European Physical Journal Plus, 126, 55

\bibitem[{{Dejonghe}(1987)}]{1987MNRAS.224...13D}
{Dejonghe}, H. 1987, \mnras, 224, 13

\bibitem[{{Di Carlo} {et~al.}(2021){Di Carlo}, {Mapelli}, {Pasquato},
  {Rastello}, {Ballone}, {Dall'Amico}, {Giacobbo}, {Iorio}, {Spera},
  {Torniamenti}, \& {Haardt}}]{2021MNRAS.507.5132D}
{Di Carlo}, U.~N., {Mapelli}, M., {Pasquato}, M., {et~al.} 2021, \mnras, 507,
  5132

\bibitem[{{Di Cintio} {et~al.}(2020){Di Cintio}, {Ciotti}, \&
  {Nipoti}}]{2020IAUS..351...93D}
{Di Cintio}, P., {Ciotti}, L., \& {Nipoti}, C. 2020, in Star Clusters: From the
  Milky Way to the Early Universe, ed. A.~{Bragaglia}, M.~{Davies}, A.~{Sills},
  \& E.~{Vesperini}, Vol. 351, 93--96

\bibitem[{{Di Cintio} {et~al.}(2017){Di Cintio}, {Livi}, {Lepri}, \&
  {Ciraolo}}]{2017PhRvE..95d3203D}
{Di Cintio}, P., {Livi}, R., {Lepri}, S., \& {Ciraolo}, G. 2017, \pre, 95,
  043203

\bibitem[{{Di Cintio} {et~al.}(2023){Di Cintio}, {Pasquato}, {Barbieri},
  {Bufferand}, {Casetti}, {Ciraolo}, {di Carlo}, {Ghendrih}, {Gunn}, {Gupta},
  {Kim}, {Lepri}, {Livi}, {Simon-Petit}, {Trani}, \&
  {Yoon}}]{2023IAUS..362..134D}
{Di Cintio}, P., {Pasquato}, M., {Barbieri}, L., {et~al.} 2023, IAU Symposium,
  362, 134

\bibitem[{{Di Cintio} {et~al.}(2021){Di Cintio}, {Pasquato}, {Kim}, \&
  {Yoon}}]{2021A&A...649A..24D}
{Di Cintio}, P., {Pasquato}, M., {Kim}, H., \& {Yoon}, S.-J. 2021, \aap, 649,
  A24

\bibitem[{{Di Cintio} {et~al.}(2022){Di Cintio}, {Pasquato}, {Simon-Petit}, \&
  {Yoon}}]{2022A&A...659A..19D}
{Di Cintio}, P., {Pasquato}, M., {Simon-Petit}, A., \& {Yoon}, S.-J. 2022,
  \aap, 659, A19

\bibitem[{{Ebisuzaki} {et~al.}(2001){Ebisuzaki}, {Makino}, {Tsuru}, {Funato},
  {Portegies Zwart}, {Hut}, {McMillan}, {Matsushita}, {Matsumoto}, \&
  {Kawabe}}]{2001ApJ...562L..19E}
{Ebisuzaki}, T., {Makino}, J., {Tsuru}, T.~G., {et~al.} 2001, \apjl, 562, L19

\bibitem[{{Freitag} \& {Benz}(2001)}]{2001A&A...375..711F}
{Freitag}, M. \& {Benz}, W. 2001, \aap, 375, 711

\bibitem[{{Gieles} \& {Zocchi}(2015)}]{2015MNRAS.454..576G}
{Gieles}, M. \& {Zocchi}, A. 2015, \mnras, 454, 576

\bibitem[{{Giersz}(2006)}]{2006MNRAS.371..484G}
{Giersz}, M. 2006, \mnras, 371, 484

\bibitem[{{Giersz} {et~al.}(2013){Giersz}, {Heggie}, {Hurley}, \&
  {Hypki}}]{2013MNRAS.431.2184G}
{Giersz}, M., {Heggie}, D.~C., {Hurley}, J.~R., \& {Hypki}, A. 2013, \mnras,
  431, 2184

\bibitem[{{Greene} {et~al.}(2020){Greene}, {Strader}, \&
  {Ho}}]{2020ARA&A..58..257G}
{Greene}, J.~E., {Strader}, J., \& {Ho}, L.~C. 2020, \araa, 58, 257

\bibitem[{{Greif}(2015)}]{2015ComAC...2....3G}
{Greif}, T.~H. 2015, Computational Astrophysics and Cosmology, 2, 3

\bibitem[{{Heggie}(2011)}]{2011BASI...39...69H}
{Heggie}, D.~C. 2011, Bulletin of the Astronomical Society of India, 39, 69

\bibitem[{{Holley-Bockelmann} {et~al.}(2008){Holley-Bockelmann},
  {G{\"u}ltekin}, {Shoemaker}, \& {Yunes}}]{2008ApJ...686..829H}
{Holley-Bockelmann}, K., {G{\"u}ltekin}, K., {Shoemaker}, D., \& {Yunes}, N.
  2008, \apj, 686, 829

\bibitem[{{Holtsmark}(1919)}]{1919AnP...363..577H}
{Holtsmark}, J. 1919, Annalen der Physik, 363, 577

\bibitem[{{Hurley} {et~al.}(2005){Hurley}, {Pols}, {Aarseth}, \&
  {Tout}}]{2005MNRAS.363..293H}
{Hurley}, J.~R., {Pols}, O.~R., {Aarseth}, S.~J., \& {Tout}, C.~A. 2005,
  \mnras, 363, 293

\bibitem[{{Hypki} \& {Giersz}(2013)}]{2013MNRAS.429.1221H}
{Hypki}, A. \& {Giersz}, M. 2013, \mnras, 429, 1221

\bibitem[{{Inayoshi} {et~al.}(2020){Inayoshi}, {Visbal}, \&
  {Haiman}}]{2020ARA&A..58...27I}
{Inayoshi}, K., {Visbal}, E., \& {Haiman}, Z. 2020, \araa, 58, 27

\bibitem[{{Kamlah} {et~al.}(2021){Kamlah}, {Leveque}, {Spurzem}, {Arca Sedda},
  {Askar}, {Banerjee}, {Berczik}, {Giersz}, {Hurley}, {Belloni},
  {K{\"u}hmichel}, \& {Wang}}]{2021arXiv210508067K}
{Kamlah}, A.~W.~H., {Leveque}, A., {Spurzem}, R., {et~al.} 2021, arXiv
  e-prints, arXiv:2105.08067

\bibitem[{{Kim} {et~al.}(2008){Kim}, {Yoon}, {Lee}, \&
  {Spurzem}}]{2008MNRAS.383....2K}
{Kim}, E., {Yoon}, I., {Lee}, H.~M., \& {Spurzem}, R. 2008, \mnras, 383, 2

\bibitem[{{Li}(2022)}]{2022arXiv220811894L}
{Li}, G.-P. 2022, arXiv e-prints, arXiv:2208.11894

\bibitem[{{Lodato} \& {Natarajan}(2006)}]{2006MNRAS.371.1813L}
{Lodato}, G. \& {Natarajan}, P. 2006, \mnras, 371, 1813

\bibitem[{{Loeb} \& {Rasio}(1994)}]{1994ApJ...432...52L}
{Loeb}, A. \& {Rasio}, F.~A. 1994, \apj, 432, 52

\bibitem[{{Londrillo} \& {Messina}(1990)}]{1990MNRAS.242..595L}
{Londrillo}, P. \& {Messina}, A. 1990, \mnras, 242, 595

\bibitem[{{Merritt}(1985)}]{1985AJ.....90.1027M}
{Merritt}, D. 1985, AJ, 90, 1027

\bibitem[{{Merritt}(2004)}]{2004cbhg.symp..263M}
{Merritt}, D. 2004, in Coevolution of Black Holes and Galaxies, ed. L.~C. {Ho},
  263

\bibitem[{{Mezcua}(2017)}]{2017IJMPD..2630021M}
{Mezcua}, M. 2017, International Journal of Modern Physics D, 26, 1730021

\bibitem[{{Miller} \& {Hamilton}(2002)}]{2002MNRAS.330..232C}
{Miller}, M.~C. \& {Hamilton}, D.~P. 2002, \mnras, 330, 232

\bibitem[{{Osipkov}(1979)}]{1979SvAL....5...42O}
{Osipkov}, L.~P. 1979, Soviet Astronomy Letters, 5, 42

\bibitem[{{Pacucci} \& {Loeb}(2022)}]{2022MNRAS.509.1885P}
{Pacucci}, F. \& {Loeb}, A. 2022, \mnras, 509, 1885

\bibitem[{{Pasquato} \& {Di Cintio}(2020)}]{2020A&A...640A..79P}
{Pasquato}, M. \& {Di Cintio}, P. 2020, \aap, 640, A79

\bibitem[{{Peebles}(1972)}]{1972ApJ...178..371P}
{Peebles}, P.~J.~E. 1972, \apj, 178, 371

\bibitem[{{Plummer}(1911)}]{1911MNRAS..71..460P}
{Plummer}, H.~C. 1911, \mnras, 71, 460

\bibitem[{{Portegies Zwart} {et~al.}(2004){Portegies Zwart}, {Baumgardt},
  {Hut}, {Makino}, \& {McMillan}}]{2004Natur.428..724P}
{Portegies Zwart}, S.~F., {Baumgardt}, H., {Hut}, P., {Makino}, J., \&
  {McMillan}, S.~L.~W. 2004, \nat, 428, 724

\bibitem[{{Rizzuto} {et~al.}(2023){Rizzuto}, {Naab}, {Rantala}, {Johansson},
  {Ostriker}, {Stone}, {Liao}, \& {Irodotou}}]{2023MNRAS.tmp..730R}
{Rizzuto}, F.~P., {Naab}, T., {Rantala}, A., {et~al.} 2023, \mnras
  [\eprint[arXiv]{2211.13320}]

\bibitem[{{Rizzuto} {et~al.}(2021){Rizzuto}, {Naab}, {Spurzem}, {Giersz},
  {Ostriker}, {Stone}, {Wang}, {Berczik}, \& {Rampp}}]{2021MNRAS.501.5257R}
{Rizzuto}, F.~P., {Naab}, T., {Spurzem}, R., {et~al.} 2021, \mnras, 501, 5257

\bibitem[{{Salpeter}(1955)}]{1955ApJ...121..161S}
{Salpeter}, E.~E. 1955, \apj, 121, 161

\bibitem[{{Sollima} \& {Ferraro}(2019)}]{2019MNRAS.483.1523S}
{Sollima}, A. \& {Ferraro}, F.~R. 2019, \mnras, 483, 1523

\bibitem[{{Sollima} \& {Mastrobuono Battisti}(2014)}]{2014MNRAS.443.3513S}
{Sollima}, A. \& {Mastrobuono Battisti}, A. 2014, \mnras, 443, 3513

\bibitem[{{Takahashi} \& {Portegies Zwart}(2000)}]{2000ApJ...535..759T}
{Takahashi}, K. \& {Portegies Zwart}, S.~F. 2000, \apj, 535, 759

\bibitem[{{Torniamenti} {et~al.}(2022){Torniamenti}, {Pasquato}, {Di Cintio},
  {Ballone}, {Iorio}, {Artale}, \& {Mapelli}}]{2022MNRAS.510.2097T}
{Torniamenti}, S., {Pasquato}, M., {Di Cintio}, P., {et~al.} 2022, \mnras, 510,
  2097

\bibitem[{{Tremou} {et~al.}(2018){Tremou}, {Strader}, {Chomiuk}, {Shishkovsky},
  {Maccarone}, {Miller-Jones}, {Tudor}, {Heinke}, {Sivakoff}, {Seth}, \&
  {Noyola}}]{2018ApJ...862...16T}
{Tremou}, E., {Strader}, J., {Chomiuk}, L., {et~al.} 2018, \apj, 862, 16

\bibitem[{{Vasiliev}(2015)}]{2015MNRAS.446.3150V}
{Vasiliev}, E. 2015, \mnras, 446, 3150

\bibitem[{{Volonteri} {et~al.}(2008){Volonteri}, {Lodato}, \&
  {Natarajan}}]{2008MNRAS.383.1079V}
{Volonteri}, M., {Lodato}, G., \& {Natarajan}, P. 2008, \mnras, 383, 1079

\bibitem[{{Wang}(2020)}]{2020MNRAS.491.2413W}
{Wang}, L. 2020, \mnras, 491, 2413

\bibitem[{{Weatherford} {et~al.}(2022){Weatherford}, {K{\i}ro{\u{g}}lu},
  {Fragione}, {Chatterjee}, {Kremer}, \& {Rasio}}]{2022arXiv221116523W}
{Weatherford}, N.~C., {K{\i}ro{\u{g}}lu}, F., {Fragione}, G., {et~al.} 2022,
  arXiv e-prints, arXiv:2211.16523

\bibitem[{{Webb} {et~al.}(2022){Webb}, {Hunt}, \& {Bovy}}]{2022arXiv221206847W}
{Webb}, J.~J., {Hunt}, J. A.~S., \& {Bovy}, J. 2022, arXiv e-prints,
  arXiv:2212.06847

\bibitem[{{Yoshida} {et~al.}(2006){Yoshida}, {Omukai}, {Hernquist}, \&
  {Abel}}]{2006ApJ...652....6Y}
{Yoshida}, N., {Omukai}, K., {Hernquist}, L., \& {Abel}, T. 2006, \apj, 652, 6

\bibitem[{{Zocchi} {et~al.}(2016){Zocchi}, {Gieles}, \&
  {H{\'e}nault-Brunet}}]{2016IAUS..312..197Z}
{Zocchi}, A., {Gieles}, M., \& {H{\'e}nault-Brunet}, V. 2016, in Star Clusters
  and Black Holes in Galaxies across Cosmic Time, ed. Y.~{Meiron}, S.~{Li},
  F.~K. {Liu}, \& R.~{Spurzem}, Vol. 312, 197--200

\end{thebibliography}
\end{document}